\documentclass[pra, aps, 10pt, twocolumn, floatfix,superscriptaddress]{revtex4-1} %%

\usepackage {graphicx}
\usepackage {amssymb}
\usepackage {amsmath}
\usepackage {verbatim}
\usepackage{color}
\usepackage[normalem]{ulem}
\usepackage[dvipsnames]{xcolor}

\begin{document}

\title{Tailoring the waveshape of optical unipolar pulses in a multi-level resonant medium} 

%  Coherent control of a multi-level resonant medium by subcycle pulses

\author{Anton Pakhomov}

\affiliation{St. Petersburg State University, Universitetskaya nab. 7/9, St. Petersburg 199034, Russia}

\author{Nikolay Rosanov} 

\affiliation{St. Petersburg State University, Universitetskaya nab. 7/9, St. Petersburg 199034, Russia}

\affiliation{Ioffe Institute, Politekhnicheskaya str. 26, St. Petersburg 194021, Russia}

\author{Mikhail Arkhipov}

\affiliation{St. Petersburg State University, Universitetskaya nab. 7/9, St. Petersburg 199034, Russia}

\author{Rostislav Arkhipov}

\affiliation{St. Petersburg State University, Universitetskaya nab. 7/9, St. Petersburg 199034, Russia}

\affiliation{Ioffe Institute, Politekhnicheskaya str. 26, St. Petersburg 194021, Russia}

\begin{abstract}
We theoretically demonstrate the possibility to tune the temporal waveform of optical unipolar pulses upon their coherent interaction with a multi-level resonant medium. This is achieved through the coherent control of the response of a multi-level resonant medium by means of half-cycle unipolar  or quasi-unipolar pulses. We show that despite the ultrabroad spectrum of half-cycle pulses one can efficiently steer the induced medium polarization through the proper choice of the parameters of the excitation pulses. As the result, producing of unipolar optical pulses of varying profile, like rectangular or triangular ones, from an extended layer of a multi-layer medium was obtained when excited by a pair of half-cycle pulses. Besides, we find out that the response of a multi-level medium for the amplitude of the driving pulses below a certain threshold can be quantitatively well approximated by the two-level model. 
\end{abstract}

\maketitle

\section{Introduction}

Over recent years significant progress has been achieved in generation of ultra-short pulses up to few-cycle and subcycle duration in the optical range \cite{Krausz, Keller, Chini, Mourou_Nobel, Biegert, Midorikawa, Mondal, Xue_2022}. As the result, the possibilities to observe and even control different ultrafast processes in matter has become available \cite{Calegari, Ramasesha, Hassan, HeSun, Hui}.

When dealing with subcycle pulses, the following physical quantity known as the electric pulse area turns out to be of crucial importance
 \cite{arkhipov2020unipolar, Arkhipov_2022_LPL}:
\begin{equation}
\vec S_E (\vec r) = \int_{-\infty}^{+\infty} \vec E(\vec r,t) dt,
\label{area}
\end{equation}
with the electric field vector $\vec E(\vec r,t)$. The pulses with the non-zero values of the electric pulse area $\vec S_E (\vec r)$ are called unipolar ones. The most prominent property of unipolar pulses is the transfer of mechanical momentum to charged particles and thus their strong acceleration \cite{arkhipov2020unipolar, arkhipov2023unipolar}. This property naturally play a key role when driving free charges, but can be also expected to strongly influence the interaction of unipolar pulses with bound charges in resonant media.

It is commonly believed that subcycle pulses can not be applied to efficiently control the resonant media because of their extremely broad spectrum. Instead, a relatively long multi-cycle pulse is thought to be needed to excite the medium transition, so that the pulse carrier frequency is equal or at least close to the transition frequency. The ultra-broad spectrum of subcycle pulses indeed makes the interaction with resonant media strongly non-resonant.

However, when the driving pulse duration is reduced to be less than the periods of the resonant transitions in the medium the key features of the light-matter interaction get completely altered \cite{arkhipov2020unipolar,arkhipov2023unipolar}. In particular, the efficiency of the medium excitation in this case is primarily determined by the value of the electric pulse area Eq.~\eqref{area} \cite{Dimitrovski_PRA, Dimitrovski_PRL, Arkhipov2019_OL}. This electric area also has a characteristic "scale", which determines the efficiency of subcycle pulse  impact on various quantum systems \cite{Arkhipov_JL_2021, Tumakov_PRA}. As the result, unipolar pulses turn out to excite either electronic levels in different quantum systems \cite{Arkhipov2019_OL, Tumakov_PRA} or rovibrational states in polar molecules \cite{Rotator_LPL, Pakhomov_2022} much better than bipolar pulses. Still these earlier studies were only done for certain simplified models of resonant media and have not addressed the propagation issues.

So far most studies of the propagation of pulses up to subcycle duration were performed using standard two-level approximation \cite{Bullough, belenov93, Kalosha, Leblond_Sazonov, Rosanov_MDB, Song, Novitsky, Leblond, Leblond2019, Sazonov_UP_1}. At the same time the validity of the two-level model clearly becomes in question when dealing with subcycle pulses. Therefore more detailed investigation is needed to find out the applicability limits of few-level models to correctly describe the subcycle pulse interaction with real-world media.

Some works have addressed the interaction and propagation dynamics of subcycle and particularly unipolar pulses with more general multi-layer resonant media. For example, the propagation of subcycle pulses in a system of multi-level atoms with a fast irreversible relaxation of the induced dipole moment, but a slow relaxation of the populations of quantum states was considered in Ref.~\cite{Sazonov_UP_2}. Under these assumptions, the formation of soliton-like unipolar objects in a non-equilibrium multi-level medium was predicted. Another case, when a stable unipolar solitons arise in a multi-level medium with one common level, was shown in Ref.~\cite{Parkhomenko}.
In Ref.~\cite{Leblond_PRA} the formation
of robust breather-type solutions was numerically obtained in a multi-level medium with several independent transitions. However, the studies on the interaction of subcycle pulses with resonant media are still poor and the understanding of the respective features still needs to be improved.

In this paper we investigate the interaction of half-cycle unipolar or quasi-unipolar pulses with a multi-level resonant medium, namely an alkali metal (sodium). We demonstrate that half-cycle pulses shorter in duration than the periods of the medium resonant transitions can still efficiently control the induced medium polarization. Using such coherent control, we demonstrate how the field emitted by a medium layer can be tailored to yield unipolar pulses of variable temporal profile. Specifically, we obtain unipolar optical pulses of different unusual waveforms, like rectangular- or triangular-shaped, when a thick medium layer possesses an inhomogeneous spatial density distribution. Besides, we examine in detail the contribution of the higher levels into the pulse-matter interaction and show that for the driving pulse strength below a certain threshold the two-level model can well describe the response of a multi-level medium when driven by half-cycle pulses.

It is important to emphasize that, to the best of our knowledge, there has been a lack of methods suggested so far for the generation of unipolar and quasi-unipolar subcycle pulses of tunable waveform in the optical range. Earlier studies were mainly concerned with the generation of half-cycle unipolar terahertz pulses, e.g. via optical rectification \cite{Citrin}, the transition radiation of a laser-produced relativistic electron bunch from a thin high-conductivity metal foil \cite{Kuratov} or through the amplification of an initially bipolar single-cycle THz pulse in a non-equilibrium plasma channel \cite{bogatskaya2021new, Bogatskaya, Bogatskaya_2}. The unipolar THz electromagnetic precursors of plateau-like shape were experimentally observed in an electro-optical GaP crystal \cite{Efimenko, Tsarev, Ilyakov}. Unipolar THz pulses of more complex waveshapes were obtained upon coherent control of a Raman-active medium in Ref.~\cite{Pakhomov_PRA_2022}. Experimental determination of the unipolarity of THz pulses was performed in Ref.~\cite{Arkhipovunipolar}.

However, the works on the waveform control of optical unipolar pulses are still lacking and were only focused on the production of half-cycle attosecond unipolar or quasi-unipolar pulses, e.g. through the optical attosecond pulse synthesis \cite{Hassan}, excitation of a foil target by intense femtosecond pulses \cite{Wu, Xu, Eliasson, Eliasson_2, Eliasson_3} or the cascaded processes in plasma \cite{Shou_Mourou}.  Meanwhile, subcycle pulses of different non-harmonic shapes in the optical range can be in high demand, e.g., for the coherent control of various ultrafast processes in matter, implementing logical operations or ultrafast controlling  qubit states \cite{Bastrakova, Bastrakova2}.

The paper is organized as follows. In Section II we present our numerical model and describe the considered multi-level resonant medium, namely the alkali metal (sodium). In Section III we investigate the coherent control of a single multi-level atom by a sequence of subcycle unipolar and bipolar pulses and show how the efficient control of the medium response can be implemented. In Section IV we consider the response of a thick medium layer driven by subcycle pulses and demonstrate the emission of both isolated unipolar half-cycle pulses in the optical range and unipolar pulses of more complex non-harmonic waveforms. Finally, we provide concluding remarks and paper summary in Section V.

\section{Model}

To describe the optical response of a multi-level resonant medium, we numerically solve dynamic equations for the expansion amplitudes $a_k(t)$ of the wave function $\psi(\vec r, t)$ of the medium over the eigen functions $\psi_k(\vec r)$:
\begin{eqnarray}
\nonumber
\psi(\vec r, t) = \sum_{k=1}^{N'} a_k(t) \psi_k(\vec r) e^{-\frac{iE_kt}{\hbar}}, 
\label{wavefunction}
\end{eqnarray}
which are given as follows \cite{Yariv}:
\begin{eqnarray}
\nonumber
\dot a_k(t) &=& \frac{i}{\hbar} \sum_{m=1}^{N'} d_{km} \ a_m(t) \ E(t) \ e^{i\omega_{km} t}, \\
\omega_{km} &=& \frac{E_k - E_m}{\hbar}, 
\label{Bloch}
\end{eqnarray}
where $N'$ is the total number of levels, $d_{km}$ is the transition dipole moment between levels $k$ and $m$, $\omega_{km}$ is the frequency of the corresponding transition and $E_k$ is the energy value in the respective eigenstate. Since we are dealing with subcycle pulses, we fully neglect the relaxation terms in Eq.~\eqref{Bloch}, as the typical lifetimes of the excited states are many orders of magnitude larger than the oscillations periods in the optical range.

We consider the excitation of the resonant medium Eq.~\eqref{Bloch} by a pair of half-cycle pulses,  having for simplicity the Gaussian profile:
\begin{eqnarray}
E(t) &=&  E_0 e^{-t^2/\tau_p^2} + E_0 e^{-(t - \Delta)^2/\tau_p^2}, 
\label{Et}
\end{eqnarray}
where $\tau_p$ is the duration of excitation pulses, $E_0$ is the electric field amplitude and $\Delta$ is the pulse-to-pulse delay. Since the medium coherence only exists over the time slot limited by the coherence relaxation time $T_2$, the coherent control can only be realized if:
\begin{eqnarray}
\Delta, \tau_p \ll T_2,
\label{coherent}
\end{eqnarray}
since otherwise the medium coherence would be damped down before the arrival of the second pulse.

%%%%%%%%%%%%%%%%%%%%%%%%%%%%%%%%%%%%%%%%%%%%%%%%%%%%%
\begin{figure}[tpb]
\centering
\includegraphics[width=1.0\linewidth]{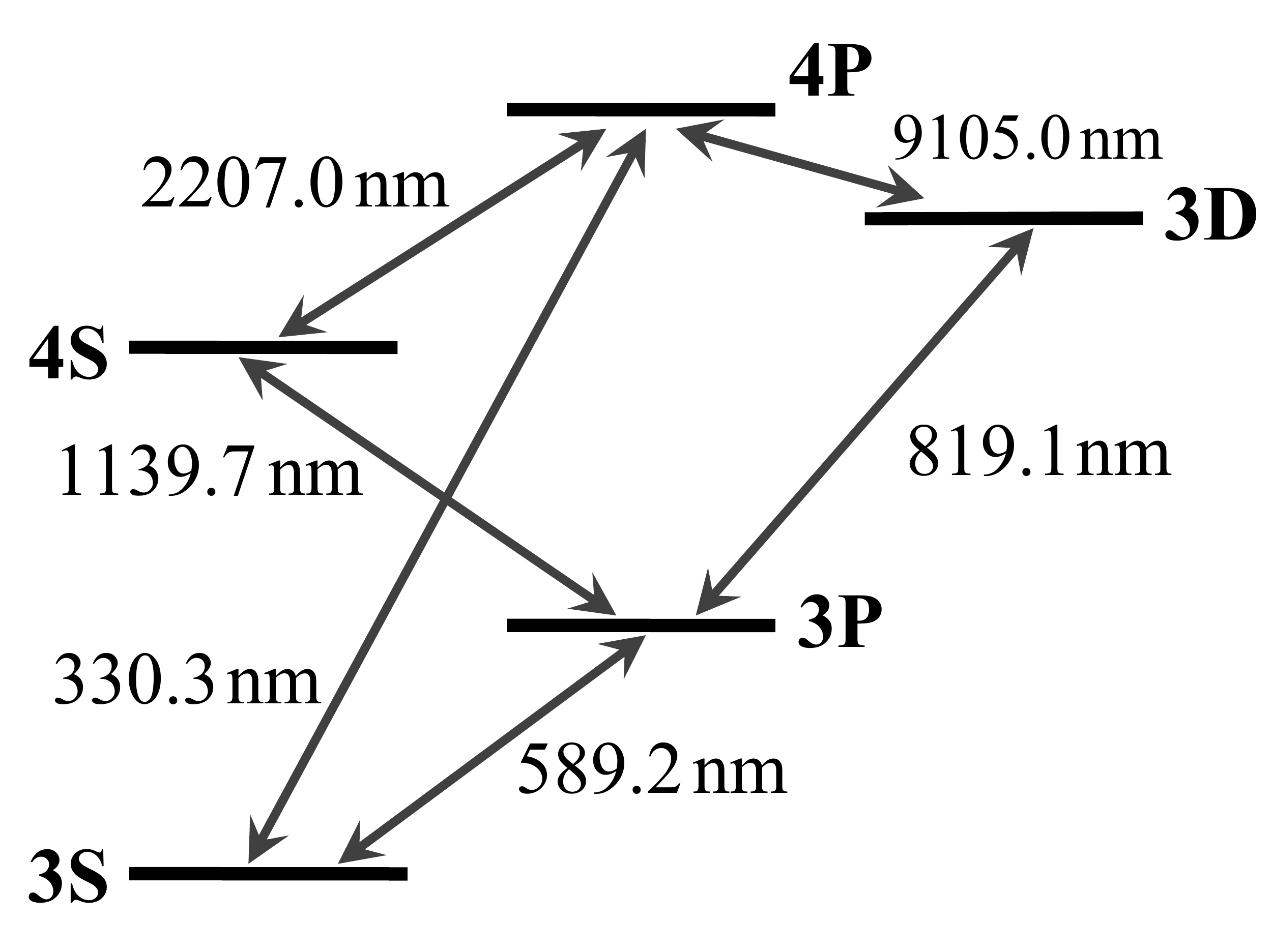}
\caption{(Color online) The energy level diagram of a sodium atom (Na) truncated to the lowest 5 levels (multiplets); arrows indicate all allowed transitions, while numbers provide the respective transition wavelengths \cite{Wiese}.}
\label{fig1}
\end{figure}
%%%%%%%%%%%%%%%%%%%%%%%%%%%%%%%%%%%%%%%%%%%%%%%%%%%%%%

We assume linearly polarized excitation pulses which are normally incident on the medium layer, so that we end up with a one-dimensional problem. The interaction of the pulses Eq.~\eqref{Et} with the medium layer can be therefore described by the scalar one-dimensional wave equation:
\begin{eqnarray}
\frac{\partial^2 E(z,t)}{\partial z^2}-\frac{1}{c^2}\frac{\partial^2 E(z,t)}{\partial t^2} = \frac{4\pi}{c^2}  \frac{\partial^2 P(z,t)}{\partial t^2}.
\label{eq_wave}
\end{eqnarray}
Besides, the one-dimensional wave equation was shown to be applicable for the propagation of ultra-short pulses, including subcycle and unipolar ones, in coaxial waveguides \cite{Rosanov2019}. To complete the numerical model one has to couple the wave equation Eq.~\eqref{eq_wave} with the  equations Eq.~\eqref{Bloch} through an explicit expression for the induced medium polarization. The corresponding expression is given as:
\begin{eqnarray}
\nonumber
P(z,t) &=& N(z) \ \sum_{k=1}^{N'} \sum_{m=1}^{N'} d_{km} \ a_k(z,t) \ a_m^*(z,t) \ e^{i\omega_{km} t} \\
&+& \ \text{c.c.}
\label{polariz}
\end{eqnarray}
with the density of resonant centers $N(z)$, which can in general be spatially-varying.

One-dimensional wave equation Eq.~\eqref{eq_wave} has the following exact solution for the emitted field  \cite{Arkhipov_OL}:
\begin{eqnarray}
E_{\mathrm{emit}}(z,t)=-\frac{2 \pi}{c} \int_{0}^{L} \frac{\partial P\Big( z',t-\frac{|z-z'|}{c}\Big )}{\partial t} dz'.
\label{Field_1D_thick} 
\end{eqnarray}
As Eq.~\eqref{Field_1D_thick} states, the emitted field from a plane medium layer is determined by the first-order temporal derivative of the induced medium polarization, unlike three-dimensional problems with the emitted field driven by the second-order temporal derivative of the induced medium polarization \cite{Landau2}. This fact is caused by the specifics of the interference of the emitted waves from different parts of a layer across its transverse dimensions.

The expression Eq.~\eqref{Field_1D_thick} therefore forces to focus on the behaviour of the temporal derivative of the medium polarization. Since the induced polarization of a multi-level medium Eq.~\eqref{polariz} contains contributions from all allowed transitions, the medium polarization should represent the superposition of multiple harmonic terms with different oscillation frequencies $\omega_{km}$. As long as we take subcycle excitation pulses with their ultra-broad spectrum and strongly non-resonant interaction with media, the relative contributions of different polarization component in Eq.~\eqref{polariz} have to be examined in details.

It seems reasonable, however, to expect the transition to the first excited state to provide the major contribution to the induced medium polarization. Therefore, as our starting point, we select the time delay between pulses $\Delta$ to equal:
\begin{eqnarray}
\Delta = \frac{T_{21}}{2} = \frac{\pi}{\omega_{21}},
\label{delta}
\end{eqnarray}
i.e. half of the period of the resonant medium oscillations at the main transition $1 \to 2$ \cite{arkhipov2020coherently}. This value is selected in such a way, that the second excitation pulse can fully stop the harmonic oscillations of the polarization at the frequency $\omega_{21}$ caused by the first pulse, similar to the idea described in Refs.~\cite{pakhomov2017all, Pakhomov_JETPL}. If the pulse-to-pulse delay differs from the value given by Eq.~\eqref{delta}, the second pulse would either speed up the polarization oscillations at the frequency $\omega_{21}$, or just partially suppress them. Thus, if one needs to obtain an isolated pulse in the medium response, the equality Eq.~\eqref{delta} has to be inevitably obeyed.

The frequency spectrum of an exciting subcycle pulse Eq.~\eqref{Et} can be easily found as:
\begin{eqnarray}
F(\omega)  \sim  \int_{-\infty}^{+\infty} E(t) e^{i \omega t} dt  =  \sqrt{\pi} E_0 \tau_p \ e^{- \omega^2 \tau_p^2 / 4},
\label{Fw}
\end{eqnarray}
i.e. steadily decreases right from the maximum at the zero frequency. According to Eq.~\eqref{delta}, it must be that:
\begin{eqnarray}
\nonumber
\tau_p  \ll  T_{21},
\label{sudden_perturb}
\end{eqnarray}
since both driving pulses in Eq.~\eqref{Et} should not overlap. Assuming that all transition frequencies in the medium are of the same order of magnitude, the exponential factor in Eq.~\eqref{Fw} yields:
\begin{eqnarray}
\nonumber
\omega_{km} \tau_p  \ll  1.
\label{sudden_perturb_2}
\end{eqnarray}
The pulse spectrum Eq.~\eqref{Fw} is therefore indeed so broad that covers the whole spectrum of the medium emission and  undergoes just slight changing across it. Still the light-medium interaction in this case is strongly non-resonant, as the pulse spectrum Eq.~\eqref{Fw} does not exhibit any features associated with any of the frequencies in the optical range.

%%%%%%%%%%%%%%%%%%%%%%%%%%%%%%%%%%%%%%%%%%%%%%%%%%%%%%%%
\begin{table}[h!]
\centering
 \begin{tabular}{||c||c|c|c|c|c||}
 \hline 
 $d_{ij}$, D &  3S  & 3P & 4S & 3D & 4P   \\
 \hline \hline
 3S & 0 & 9.06 & 0 & 0 & 0.82  \\
 \hline
 3P & 9.06 & 0 & 8.89 & 17.05 & 0  \\
 \hline
 4S & 0 & 8.89 & 0 & 0 & 20.60  \\
 \hline
 3D & 0 & 17.05 & 0 & 0 & 27.49  \\
 \hline
 4P & 0.82 & 0 & 20.60 & 27.49 & 0  \\
 \hline
\end{tabular}
\caption{The transition dipole moments for the 5 lowest levels of a sodium atom. All values are expressed in Debays (D).}
\label{table1}
\end{table}
%%%%%%%%%%%%%%%%%%%%%%%%%%%%%%%%%%%%%%%%%%%%%%%%%%%%%%%%

As a specific example of a multi-level medium we find it convenient to take an alkali metal vapour because of their relatively simple energy level structure. This level structure  basically represents a noble gas structure plus only one external electron. As a result, it is well justified to neglect all electron-electron correlations and use the single active electron approximation. Besides, the values of the coherence relaxation times in alkali metal vapours are in the nanosecond range, what ensures the condition Eq.~\eqref{coherent} for optical transitions.

Specifically, we take here the sodium atom (Na), since it has much weaker spit-orbit splitting as compared to heavier alkali-metal atoms, and this may potentially allow to avoid additional complexity in the medium response. We consider below 5 lowest energy levels of a sodium atom, while neglecting for the sake of simplicity the fine structure of levels. Namely, we focus on the following energy terms: 3S (ground state), 3P, 4S, 3D and 4P, see Fig.~\ref{fig1}. The P- and D-levels, which represent multiplets due to the spin-orbit splitting (e.g., the well known D-line for the transition 3S $\to$ 3P), we treat as single degenerate levels using standard summation rules for the respective line strengths \cite{Axner}. Besides, we also scale the tabulated values of the transition dipole moments taken from Ref.~\cite{Wiese} by the factor $\sqrt{3}$, since we assume here linearly polarized driving pulses, while the tabulated values were calculated for a non-polarized light. The resulting transition dipole moments for the considered sodium atom are listed in Table~\ref{table1}. In the next section the interaction of such a sodium atom with a sequence of subcycle unipolar pulses Eq.~\eqref{Et} will be addressed.

\section{Coherent control of a thin slice of multi-level atoms}

%%%%%%%%%%%%%%%%%%%%%%%%%%%%%%%%%%%%%%%%%%%%%%%%%%%%%
\begin{figure}[tpb]
\centering
\includegraphics[width=1.0\linewidth]{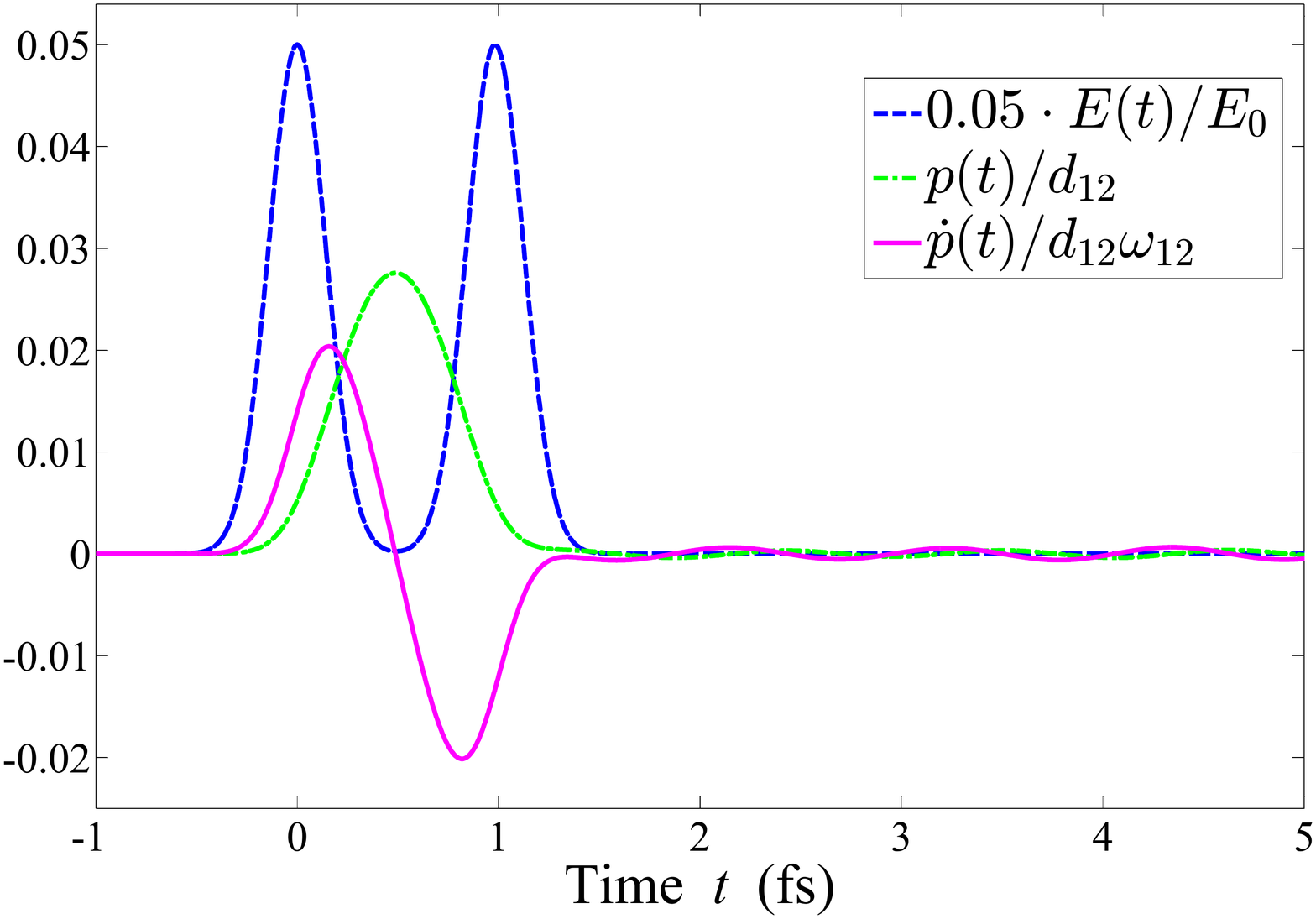}
\caption{(Color online) The induced polarization per a single atom (dash-dotted green curve) together with its temporal derivative (magenta solid curve); blue dashed line depicts the electric field of the excitation pulses of the amplitude $E_0 = 10^4$ ESU and the duration $\tau_p = 200$ as; the time delay between the excitation pulses is $\Delta = 0.98$ fs, as given by Eq.~\eqref{delta}. All quantities are plotted in dimensionless units.}
\label{fig2}
\end{figure}
%%%%%%%%%%%%%%%%%%%%%%%%%%%%%%%%%%%%%%%%%%%%%%%%%%%%%%

We begin with the excitation of an optically thin layer of the considered multi-level sodium atoms. According to Eq.~\eqref{Field_1D_thick} the emitted field in this case up to a constant factor is given by the temporal derivative of the induced polarization. Thanks to the small layer thickness, we can neglect the transformations of the excitation pulses upon their propagation across the medium layer. That is why this problem simply reduces to the response of a single  multi-level sodium atom. We proceed therefore in this section with solving the equations Eq.~\eqref{Bloch} with subcycle driving pulses for one atom and omitting the wave equation Eq.~\eqref{eq_wave} because of the negligible propagation effects.

Fig.~\ref{fig2} illustrates the obtained response of a single resonant atom upon excitation by a pair of subcycle unipolar pulses given by Eq.~\eqref{Et}. Particularly, we depict in Fig.~\ref{fig2} both the induced dipole moment of a single atom $p(t)$ as well as its temporal derivative. The pulse duration was taken $\tau_p = 200$ as, what is below the period of all resonant transitions, and the pulse amplitude was first taken as strong as $E_0 = 10^4$ ESU. One can see that the induced dipole moment to a high accuracy represents a half-sine wave, i.e. the first driving subcycle pulse induces the oscillations of the dipole moment and the second driving pulse almost completely stops the oscillations. Some residual low-amplitude oscillations of the dipole moment visible in Fig.~\ref{fig2} are caused by the contributions of other excited levels in the medium. The derivative of the induced dipole moment, i.e the emitted electric field according to Eq.~\eqref{Field_1D_thick}, forms a single sign-changing oscillation, meaning that the medium slice is to emit an almost isolated single-cycle pulse.

%%%%%%%%%%%%%%%%%%%%%%%%%%%%%%%%%%%%%%%%%%%%%%%%%%%%%
\begin{figure}[tpb]
\centering
\includegraphics[width=1.0\linewidth]{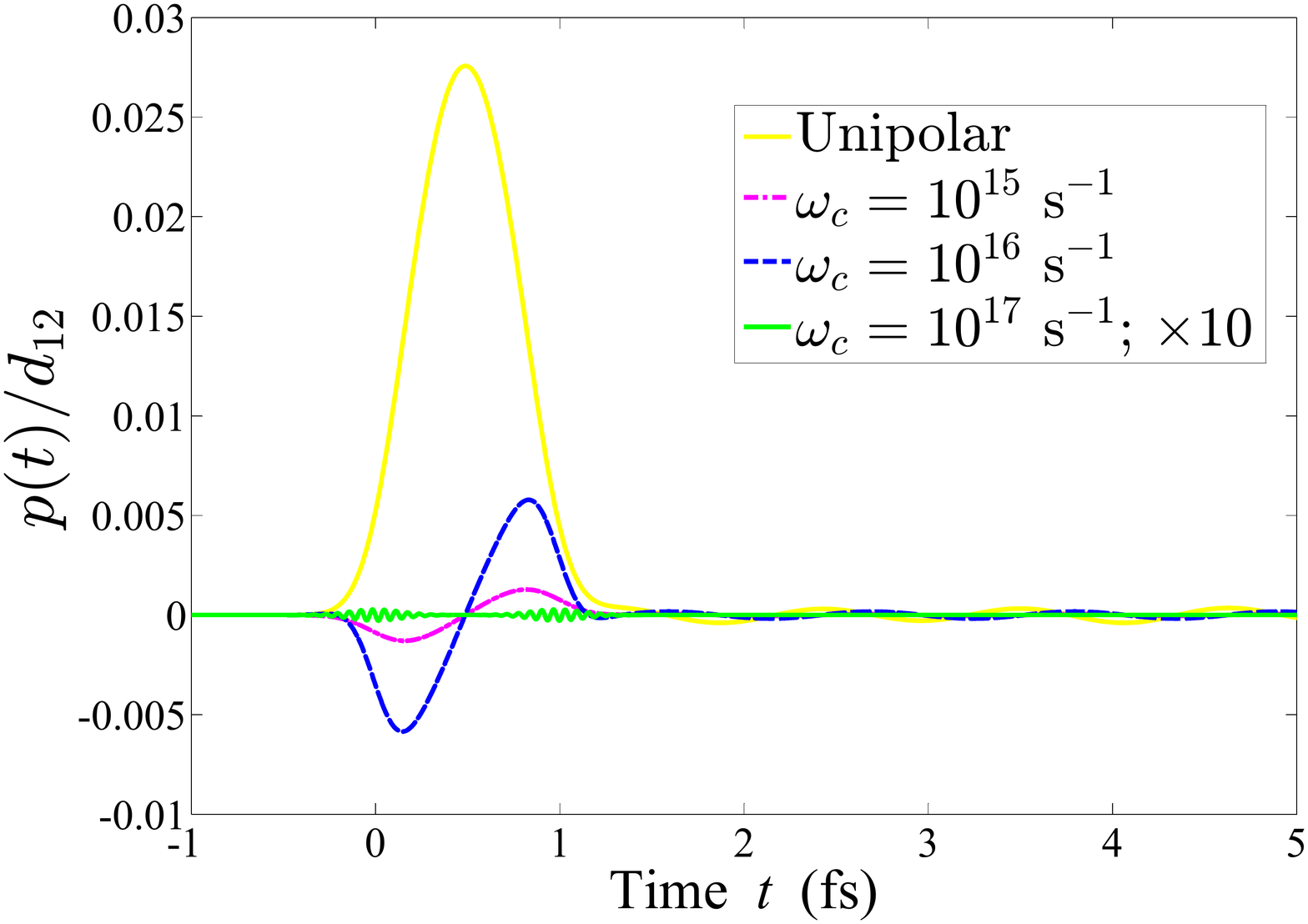}
\caption{(Color online) The induced polarization per a single atom upon the excitation by unipolar pulses Eq.~\eqref{Et} (yellow solid curve) and bipolar pulses Eq.~\eqref{Et_bip} with different carrier frequencies; the pulse duration was fixed to $\tau_p = 200$ as; all other  parameters are the same as in Fig.~\ref{fig2}. }
\label{fig3}
\end{figure}
%%%%%%%%%%%%%%%%%%%%%%%%%%%%%%%%%%%%%%%%%%%%%%%%%%%%%%

It is interesting to find out the role of the electric pulse area Eq.~\eqref{area} in the medium excitation. Therefore we also examine the case, when the same multi-level sodium atom is driven by a pair of bipolar pulses, i.e. with exactly zero electric pulse area, in the form:
\begin{eqnarray}
\nonumber
E(t) &=&  E_0 e^{-t^2/\tau_p^2} \sin \omega_c t + \\
&& E_0 e^{-(t - \Delta)^2/\tau_p^2} \sin \omega_c (t - \Delta),
\label{Et_bip}
\end{eqnarray}
where $\omega_c$ is the carrier frequency. Fig.~\ref{fig3} demonstrates the induced dipole moment of a single atom when excited by unipolar pulses Eq.~\eqref{Et} and bipolar pulses Eq.~\eqref{Et_bip} with different carrier frequencies $\omega_c$. The amplitude and the duration of the driving pulses were assumed constant.

Fig.~\ref{fig3} yields that unipolar subcycle pulses much more efficiently induce the medium polarization as compared to bipolar pulses. As the carrier frequency in Fig.~\ref{fig3} increases, the excitation pulses gradually transform from subcycle to single-cycle and then to few-cycle ones. At the same time, the more optical cycles the driving pulses contain, the weaker medium response we obtain. The seeming exception for $\omega_c = 10^{15}$ s$^{-1}$ and $\omega_c = 10^{16}$ s$^{-1}$ is attributed to the fact that pulses Eq.~\eqref{Et_bip} stay single-cycle for carrier frequencies up to $\sim 10^{16}$ s$^{-1}$. It is worth noting that for few-cycle driving pulses the temporal derivative of the induced polarization in Fig.~\ref{fig3}, i.e. the emitted field, represents also a few-cycle pulse instead of a single-cycle one in Fig.~\ref{fig2}.

The results seen in Fig.~\ref{fig3} well correlate with earlier studies of the  excitation of several model media by unipolar pulses \cite{Arkhipov2019_OL, Rotator_LPL, Tumakov_PRA, Pakhomov_2022}. In these works it was found that for pulse durations shorter than the resonant transition periods the electric pulse area drives the medium excitation. Hence, unipolar pulses provide more efficient non-resonant excitation than bipolar ones. We can state that this result stays valid for the excitation of more complex multi-level media and holds not only for the level populations, but also for the induced medium polarization.

%%%%%%%%%%%%%%%%%%%%%%%%%%%%%%%%%%%%%%%%%%%%%%%%%%%%%
\begin{figure}[tpb]
\centering
\includegraphics[width=1.0\linewidth]{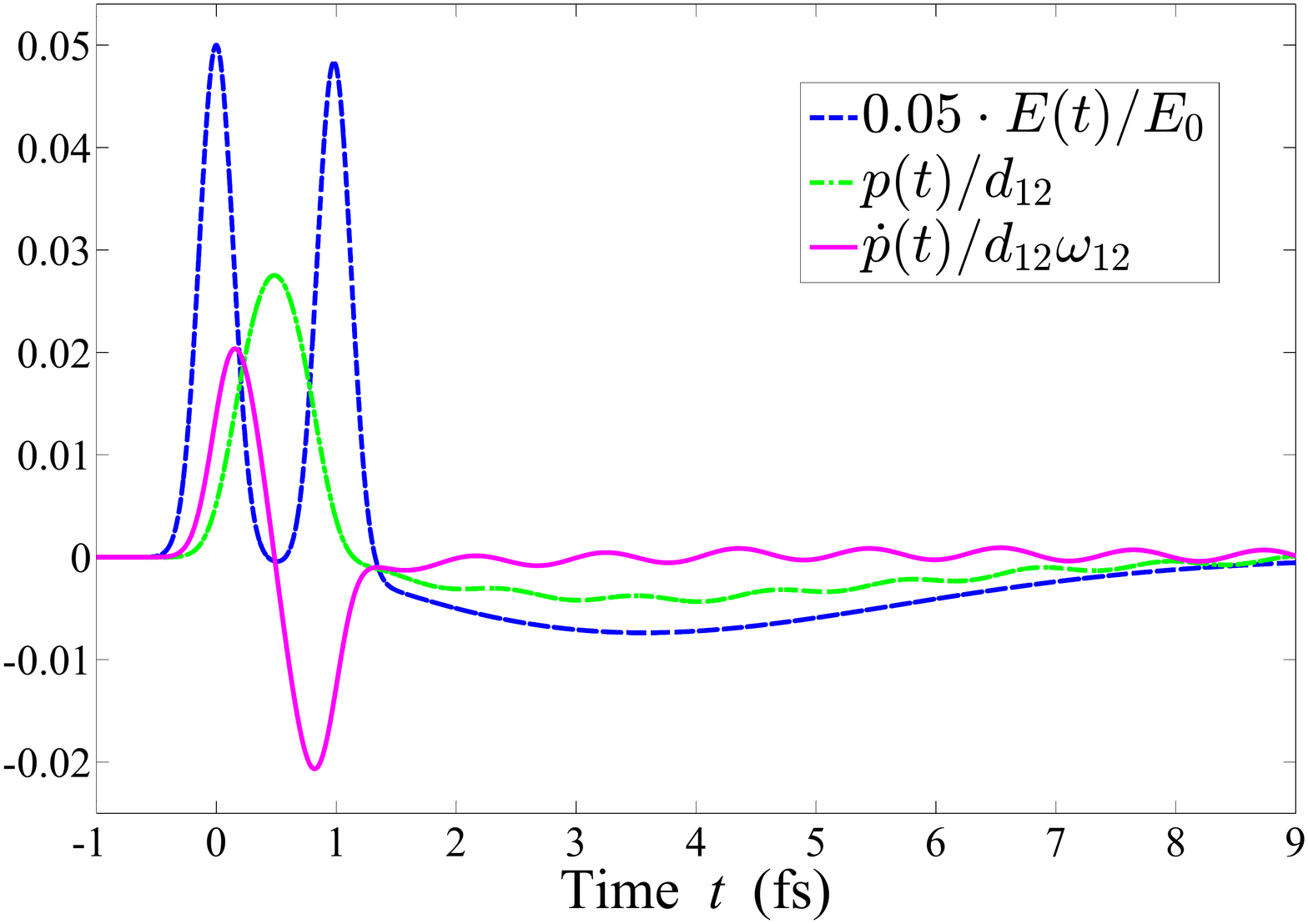}
\caption{(Color online)   The induced polarization per a single atom (dash-dotted green curve) together with its temporal derivative (magenta solid curve) upon the excitation by quasi-unipolar pulses Eq.~\eqref{Et_1}; blue dashed line depicts the electric field of the excitation pulses of the amplitude $E_0 = 10^4$ ESU, the duration $\tau_p = 200$ as and the tail width $\tau_1 = 20 \tau_p$; the time delay between the excitation pulses is $\Delta = 0.98$ fs, as given by Eq.~\eqref{delta}. All quantities are plotted in dimensionless units.}
\label{fig3-2}
\end{figure}
%%%%%%%%%%%%%%%%%%%%%%%%%%%%%%%%%%%%%%%%%%%%%%%%%%%%%%

%%%%%%%%%%%%%%%%%%%%%%%%%%%%%%%%%%%%%%%%%%%%%%%%%%%%%
\begin{figure}[tpb]
\centering
\includegraphics[width=1.0\linewidth]{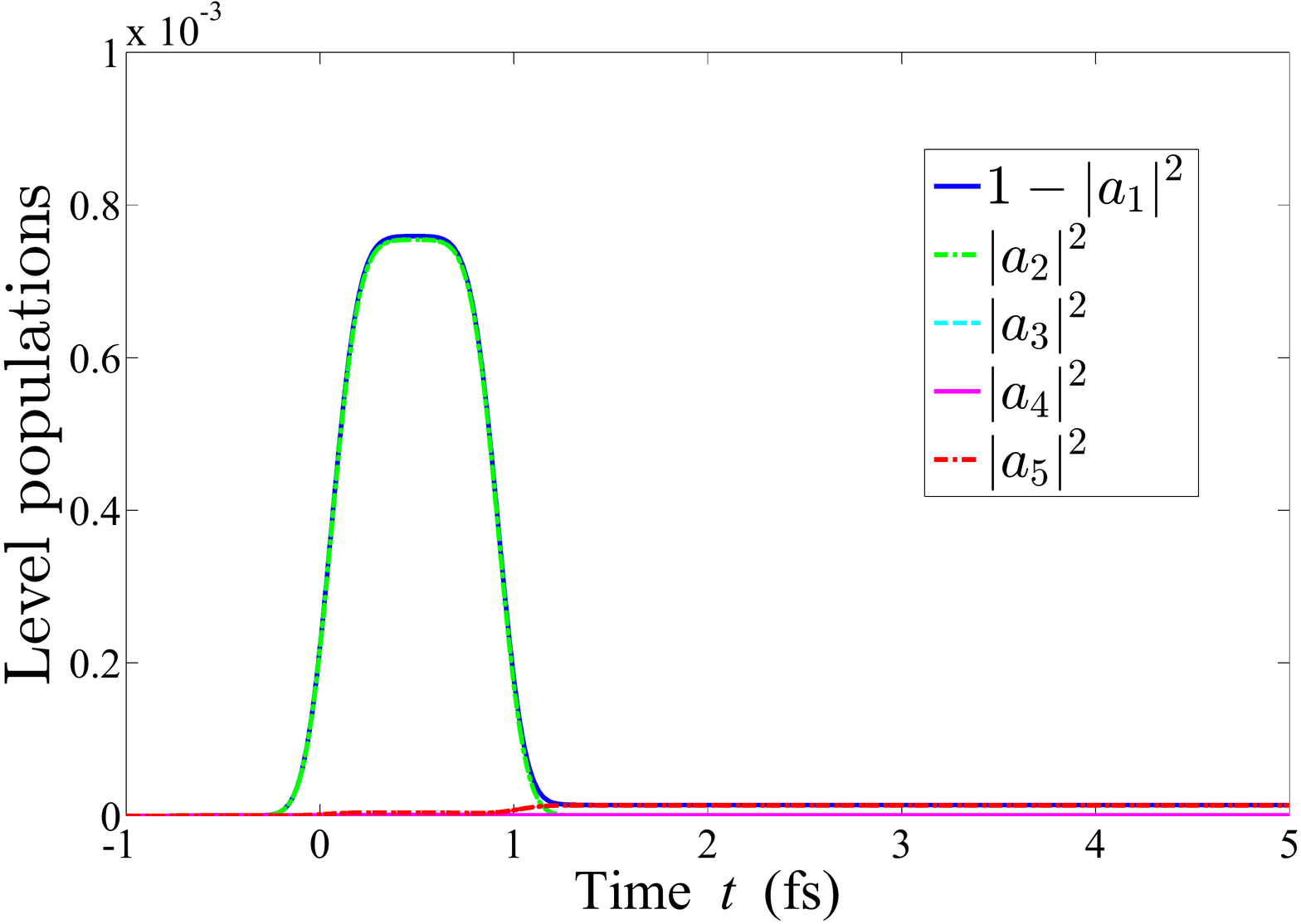}
\caption{(Color online) The populations of atom's levels vs. time; all  parameters are the same as in Fig.~\ref{fig2}. }
\label{fig4}
\end{figure}
%%%%%%%%%%%%%%%%%%%%%%%%%%%%%%%%%%%%%%%%%%%%%%%%%%%%%%

Finally, it seems important to address the condition on the exact value of the electric pulse area Eq.~\eqref{area} of the exciting pulses. In fact, obtaining exactly unipolar half-cycle pulses like Eq.~\eqref{Et} in experiments could be quite tricky. Instead, it is usually much easier to produce so-called quasi-unipolar or quasi-half-cycle pulses, containing an intense half-cycle burst of one polarity and a long weak tail of opposite polarity. As the result, the total electric pulse area Eq.~\eqref{area} equals zero, what allows to avoid a number of complex issues related to dealing with near-zero-frequency spectral components. That is why we are specifically interested to  consider the medium excitation by quasi-unipolar pulses and compare the response with the one shown in Figs.~\ref{fig2}-\ref{fig3}.

For this end instead of Eq.~\eqref{Et} we take each driving pulse in the following form:
\begin{eqnarray}
\nonumber
E(t) &=&  E_0 \Big( e^{-t^2/\tau_p^2} - \kappa  (t- \tau_p) \ e^{-(t - \tau_p)^2/\tau_1^2} \ \Theta [t - \tau_p] \Big), \\
\label{Et_1}
\end{eqnarray}
with the Heaviside step function $\Theta$ and the tail temporal width $\tau_1$. With this specific choice of the temporal profile Eq.~\eqref{Et_1} it is easy to calculate the electric pulse area Eq.~\eqref{area} as:
\begin{eqnarray}
\nonumber
S_E =  E_0 \Big( \tau_p \sqrt{\pi} - \frac{\kappa}{2} \tau_1^2 \Big).
\label{area_quasi}
\end{eqnarray}
Fixing the pulse parameters to obey the equality:
\begin{eqnarray}
\kappa =  \frac{2 \sqrt{\pi} \tau_p}{\tau_1^2}
\label{area_quasi_2}
\end{eqnarray}
we end up with the exactly zero electric pulse area. In the case of $\tau_1 \gg \tau_p$ we get a proper analytical approximation of a quasi-unipolar half-cycle pulse possessing a long low-amplitude tail. We have performed the solution of the medium equations Eq.~\eqref{Bloch} driven by a pair of quasi-half-cycle pulses Eq.~\eqref{Et_1} with the delay given by Eq.~\eqref{delta}. Our calculations show that if the tail width $\tau_1$ at least by an order of magnitude exceeds the duration of the main half-cycle burst $\tau_p$, the medium response turns out to be close to the one for exactly unipolar half-cycle pulses in 
Fig.~\ref{fig2}. An example is shown in Fig.~\ref{fig3-2} for the values $\tau_1 = 20 \tau_p$ and the parameter $\kappa$ from Eq.~\eqref{area_quasi_2}. One can see that despite the presence of the long tail of the pulses the emission of a thin medium layer still represents a single-cycle pulse like in Fig.~\ref{fig2}, while the tail only causes the slight oscillations at the trailing edge of the medium emission. Hence, based on the Figs.~\ref{fig3}-\ref{fig3-2}, we proceed below with the medium excitation by unipolar pulses, as provided by Eq.~\eqref{Et}, keeping in mind that quasi-unipolar pulses would lead to basically the same medium response.

In Fig.~\ref{fig4} we show the time trace of the populations of all medium levels for the case from Fig.~\ref{fig2}. It is well seen that only the first excited level (3P) gets noticeably populated by the first driving pulse, while the populations of higher levels stay negligibly small. After the action of the second driving pulse, though, the first excited level becomes almost completely depopulated and the largest residual population remains at the fifth level (4P), but still low enough to be disregarded (around $10^{-5}$).

\begin{figure}[tpb]
\centering
\includegraphics[width=1.0\linewidth]{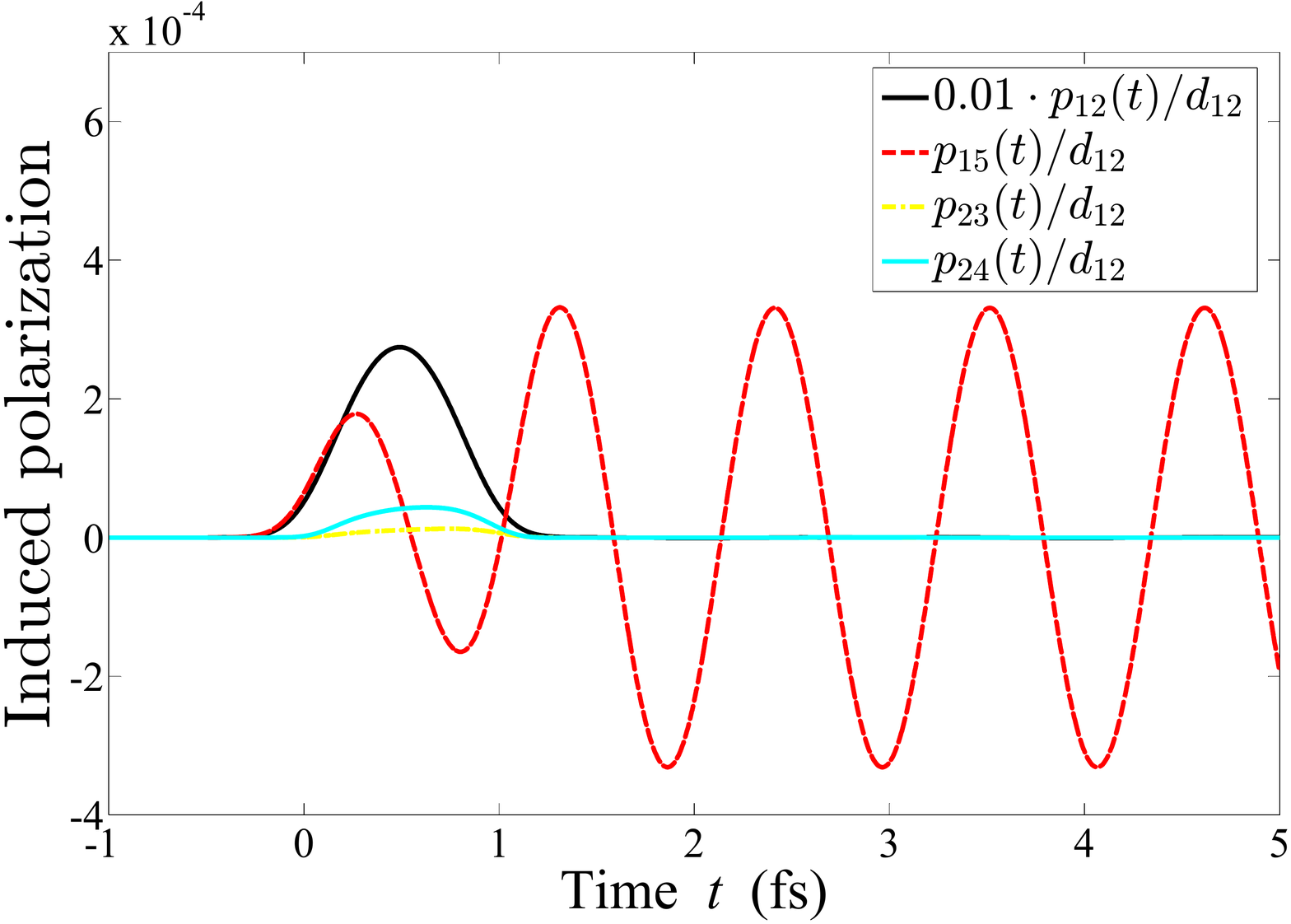}
\caption{(Color online) The strongest polarization components of a single atom vs. time; all  parameters are the same as in Fig.~\ref{fig2}. }
\label{fig5}
\end{figure}

One can therefore see that despite the ultrabroad spectrum of driving unipolar pulses the multi-level medium behaves pretty much like a two-level medium. To get a closer look at the dynamics of the induced medium polarization we plot in Fig.~\ref{fig5} separately contributions of different transitions into the induced dipole moment of a single sodium atom according to Eq.~\eqref{polariz}, so that:
\begin{eqnarray}
p_{km} = d_{km}  a_k  a_m^* \ e^{i\omega_{km} t} + \text{c.c.}
\label{p_km}
\end{eqnarray}
Only the largest polarization components are displayed in Fig.~\ref{fig5}, while the rest ones are much smaller in amplitude. As can be seen from Fig.~\ref{fig5}, the contribution of the main transition $p_{12}$ for at least two orders of magnitude exceeds those for all other possible transitions. That is why the medium response in Fig.~\ref{fig2} closely follows the behaviour predicted for a two-level medium.

\begin{figure}[tpb]
\centering
\includegraphics[width=1.0\linewidth]{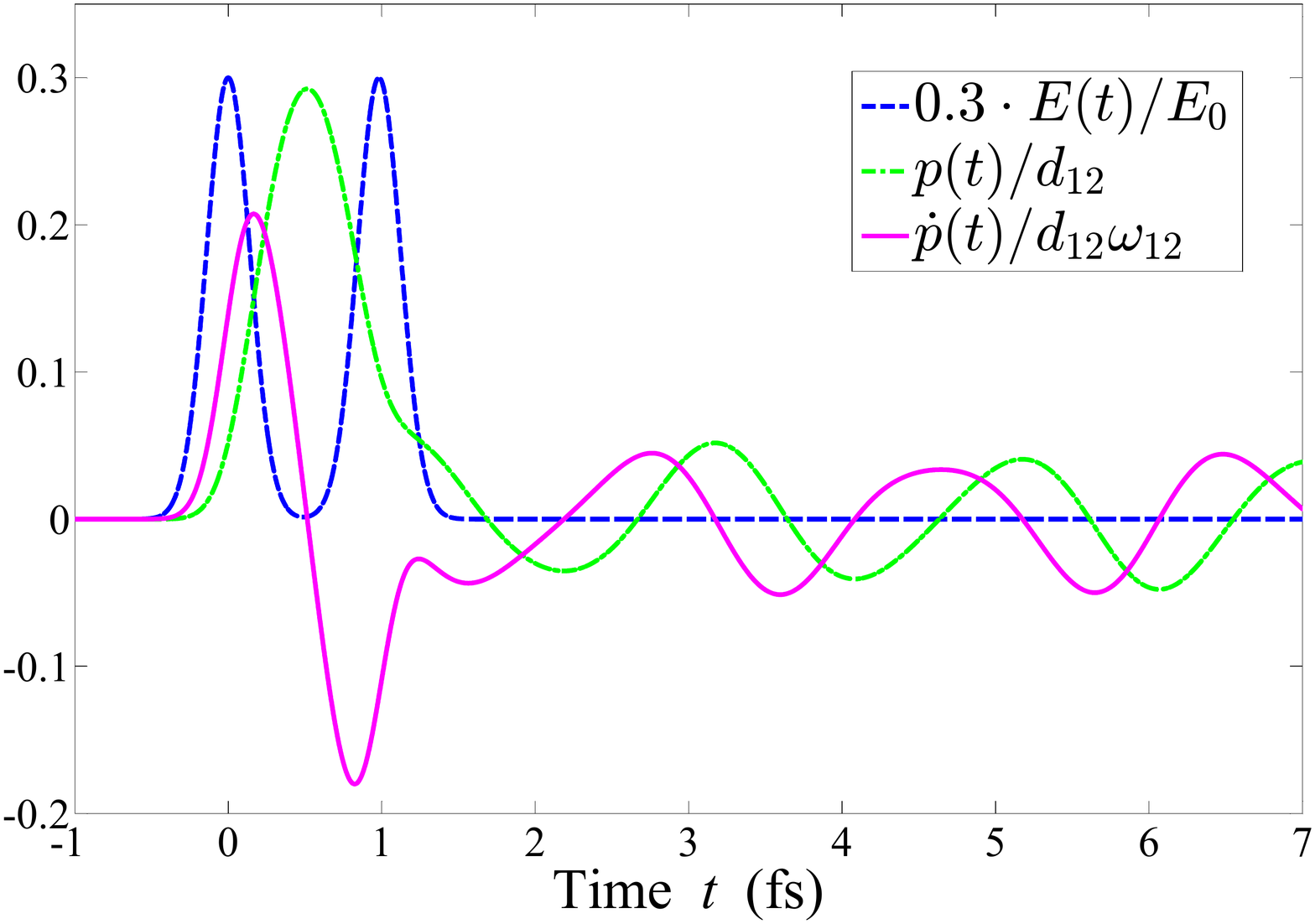}
\caption{(Color online) The induced polarization per a single atom (dashed green curve) together with its temporal derivative (magenta solid curve) and the electric field of the excitation pulses (black solid line); the pulse amplitude $E_0 = 10^5$ ESU, all other parameters are the same as in Fig.~\ref{fig2}. All quantities are plotted in dimensionless units.}
\label{fig6}
\end{figure}

However, when further increasing the field strength of the driving pulses the contributions of higher excited states become more pronounced and the discrepancy from the two-level medium's dynamics starts growing. In Fig.~\ref{fig6} we demonstrate the response of a single atom similar to shown in Fig.~\ref{fig2}, but with the pulse amplitude increased to $E_0 = 10^5$ ESU. It is evident that the residual field oscillations now become comparable in amplitude to the leading peak. They arise because the second excitation pulse can not anymore stop the polarization oscillations initiated by the first pulse due to the significant contribution of higher levels. As the result, the polarization profile in Fig.~\ref{fig6} can not be treated as a half-sine, but rather has a pronounced oscillating tail.

%%%%%%%%%%%%%%%%%%%%%%%%%%%%%%%%%%%%%%%%%%%%%%%%%%%%%
%%%%%%%%%%%%%%%%%%%%%%%%%%%%%%%%%%%%%%%%%%%%%%%%%%%%%%

\begin{figure}[tpb]
\centering
\includegraphics[width=1.0\linewidth]{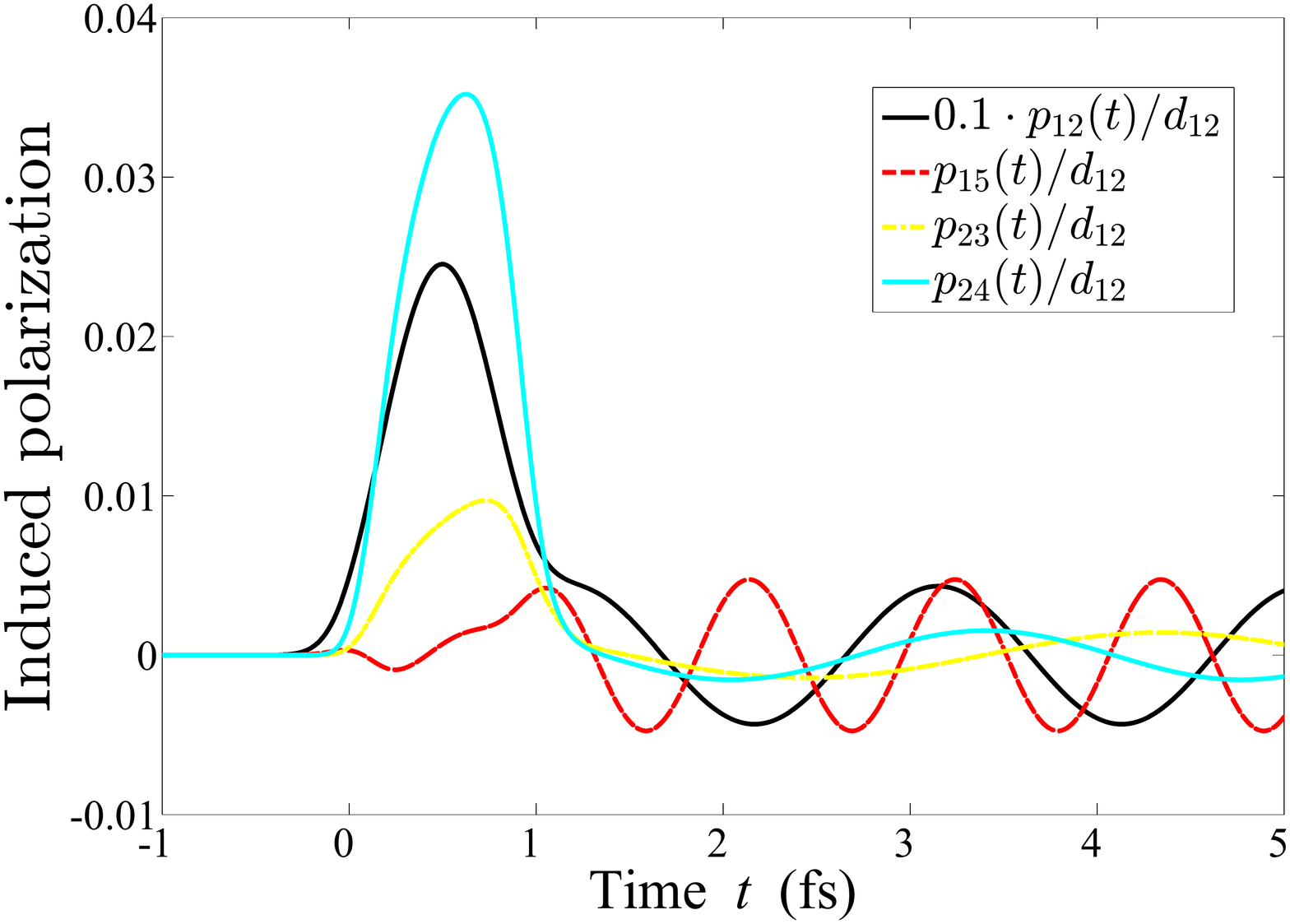}
\caption{(Color online) The polarization components of a single atom vs. time for the parameters from Fig.~\ref{fig6}. }
\label{fig7}
\end{figure}

Fig.~\ref{fig7} shows the contributions of different transitions into the induced dipole moment of a single atom for the parameters of Fig.~\ref{fig6}. It is well seen that not only the contributions of other transitions become larger, but also the component $p_{12}$ exhibits residual oscillations thanks to the stronger interplay between all considered excited levels.

Let us try to analytically calculate the threshold pulse amplitude for the applicability of the two-level approximation. For this end we make use of the perturbation theory. The perturbation theory implies the small populations of the excited states (much less than unity) and the population of the ground state to stay close to unity. We therefore assume in the equations Eq.~\eqref{Bloch} $a_1(t) \approx 1$ and obtain:
\begin{eqnarray}
a_k (t) \approx \frac{i d_{1k}}{\hbar} \int_{-\infty}^t E(t') e^{i \omega_{k1} t'} dt', \ \ k > 1.
\label{perturb_1}
\end{eqnarray}
If the driving pulse duration is chosen well below the resonant transition periods in the medium:
\begin{eqnarray}
\nonumber
\omega_{km} \tau_p \ll 1
\label{shortness}
\end{eqnarray}
for any $k, m$, then we can take under the integral sign $e^{i \omega_{k1} t'} \approx 1$ and reduce Eq.~\eqref{perturb_1} to:
\begin{eqnarray}
a_k (t) \approx \frac{i d_{1k}}{\hbar} \int_{-\infty}^t E(t') dt', \ \ k > 1.
\label{perturb_2}
\end{eqnarray}
The expansion amplitudes $a_k (t)$ right after the passage of the first driving pulse are thus given as:
\begin{eqnarray}
a_k^{(1)} \approx \frac{i d_{1k}}{\hbar} S_E, \ \ k > 1,
\label{perturb_3}
\end{eqnarray}
i.e. are fully determined by the electric pulse area Eq.~\eqref{area}. In between two driving pulses according to Eq.~\eqref{Bloch} the amplitudes $a_k (t)$ stay constant. The polarization components $p_{1k}$ at the same time change harmonically according to Eq.~\eqref{p_km} as:
\begin{eqnarray}
p_{1k} = p_{k1} = \frac{2 d^2_{1k}}{\hbar} S_E  \cos \omega_{k1} t.
\label{p_1k}
\end{eqnarray}
After the passage of the second driving pulse we find in a similar way:
\begin{eqnarray}
\nonumber
a_k^{(2)} &\approx& \frac{i d_{1k}}{\hbar} S_E \Big( 1 + e^{i \omega_{k1} \Delta} \Big) \\
&=& \frac{i d_{1k}}{\hbar} S_E \Big( 1 + e^{i \pi \frac{\omega_{k1}}{\omega_{21}} } \Big), \ \ k > 1.
\label{perturb_4}
\end{eqnarray}
In particular, we see that due to the choice of the pulse-to-pulse delay in Eq.~\eqref{delta}:
\begin{eqnarray}
a_2^{(2)} \approx 0,
\label{perturb_5}
\end{eqnarray}
i.e. the first excited level gets fully depleted by the second driving pulse. In other excited states, however, there are some small residual populations in Eq.~\eqref{perturb_4} since $\omega_{k1} \ne \omega_{21}$.

%%%%%%%%%%%%%%%%%%%%%%%%%%%%%%%%%%%%%%%%%%%%%%%%%%%%%
%%%%%%%%%%%%%%%%%%%%%%%%%%%%%%%%%%%%%%%%%%%%%%%%%%%%%%

Typically in different resonant media the transition dipole moment between the ground and the first excited states $d_{12}$ for orders of magnitude exceeds those for all other excited states. This is also true for alkali metals and specifically for sodium, as can be seen from Table~\ref{table1} and in Ref.~\cite{Wiese} for higher levels. Since the contributions of the polarization components $p_{1k}$ according to Eq.~\eqref{p_1k} are proportional to the squared dipole moments $\sim d^2_{1k}$, the contribution of the main transition $1 \to 2$ into the induced medium polarization has to be prevailing within the perturbation theory. The validity of the perturbation theory Eq.~\eqref{perturb_1} therefore assures the validity of the two-level description.

%%%%%%%%%%%%%%%%%%%%%%%%%%%%%%%%%%%%%%%%%%%%%%%%%%%%%
%%%%%%%%%%%%%%%%%%%%%%%%%%%%%%%%%%%%%%%%%%%%%%%%%%%%%%

The expressions Eqs.~\eqref{perturb_1}-\eqref{perturb_5} are valid only as long as:
\begin{eqnarray}
\nonumber
1 - | a_1 (t) | &\ll& 1, \\
| a_k (t) | &\ll& 1, \ \ \text{for} \ k > 1.
\label{perturb_6}
\end{eqnarray}
When the inequalities Eq.~\eqref{perturb_6} become invalid, higher-order correction terms of the perturbation theory have to be added. In order to calculate the second-order terms one has to insert the first-order solutions Eq.~\eqref{perturb_1} into the right-hand side of the equations Eq.~\eqref{Bloch}. For instance, for the ground state we can find:
\begin{eqnarray}
\nonumber
a_1 (t) = 1 - \frac{1}{\hbar^2} \sum_{m=2}^{N'} d^2_{1m} \int_{-\infty}^t E(t') dt' \int_{-\infty}^{t'} E(t'') dt''.  \\
\label{ground}
\end{eqnarray}
Similarly, the expression Eq.~\eqref{perturb_1} for the amplitudes of the excited levels transforms into:
\begin{eqnarray}
\nonumber
a_k (t) &=& \frac{i d_{1k}}{\hbar} \int_{-\infty}^t E(t') e^{i \omega_{k1} t'} dt' - \\
\nonumber
&& \frac{1}{\hbar^2} \sum_{m=2}^{N'} d_{km} d_{1m} \int_{-\infty}^t E(t') dt' \int_{-\infty}^{t'} E(t'') dt''. \\
\label{excited}
\end{eqnarray}
One can see from the expressions Eqs.~\eqref{ground}-\eqref{excited} above that the second-order correction terms by the order of magnitude are:
\begin{eqnarray}
\nonumber
\sim \frac{d^2_{1k}  E_0^2 \tau_p^2}{\hbar^2}.
\label{correction}
\end{eqnarray}
The two-level approximation then ceases to describe the medium response, when these second-order correction terms Eqs.~\eqref{ground}-\eqref{excited} become comparable to the first-order terms Eq.~\eqref{perturb_1}. It seems reasonable to fix the corresponding threshold value of the electric field strength of the exciting pulses $E_{\rm{thr}}$ to the following relation:
\begin{eqnarray}
\text{max}_k d_{1k} \ \frac{ E_{\rm{thr}} \tau_p}{\hbar}  \sim 0.1.
\label{criterion}
\end{eqnarray}
Taking here $d_{1k} \sim 10$ D, $\tau_p \sim 0.1$ fs, we get the following estimate for the threshold value of the field strength:
\begin{eqnarray}
E_{\text{thr}} \sim 10^5 \ \text{ESU}.
\label{threshold}
\end{eqnarray}

%%%%%%%%%%%%%%%%%%%%%%%%%%%%%%%%%%%%%%%%%%%%%%%%%%%%%
\begin{figure}[tpb]
\centering
\includegraphics[width=1.0\linewidth]{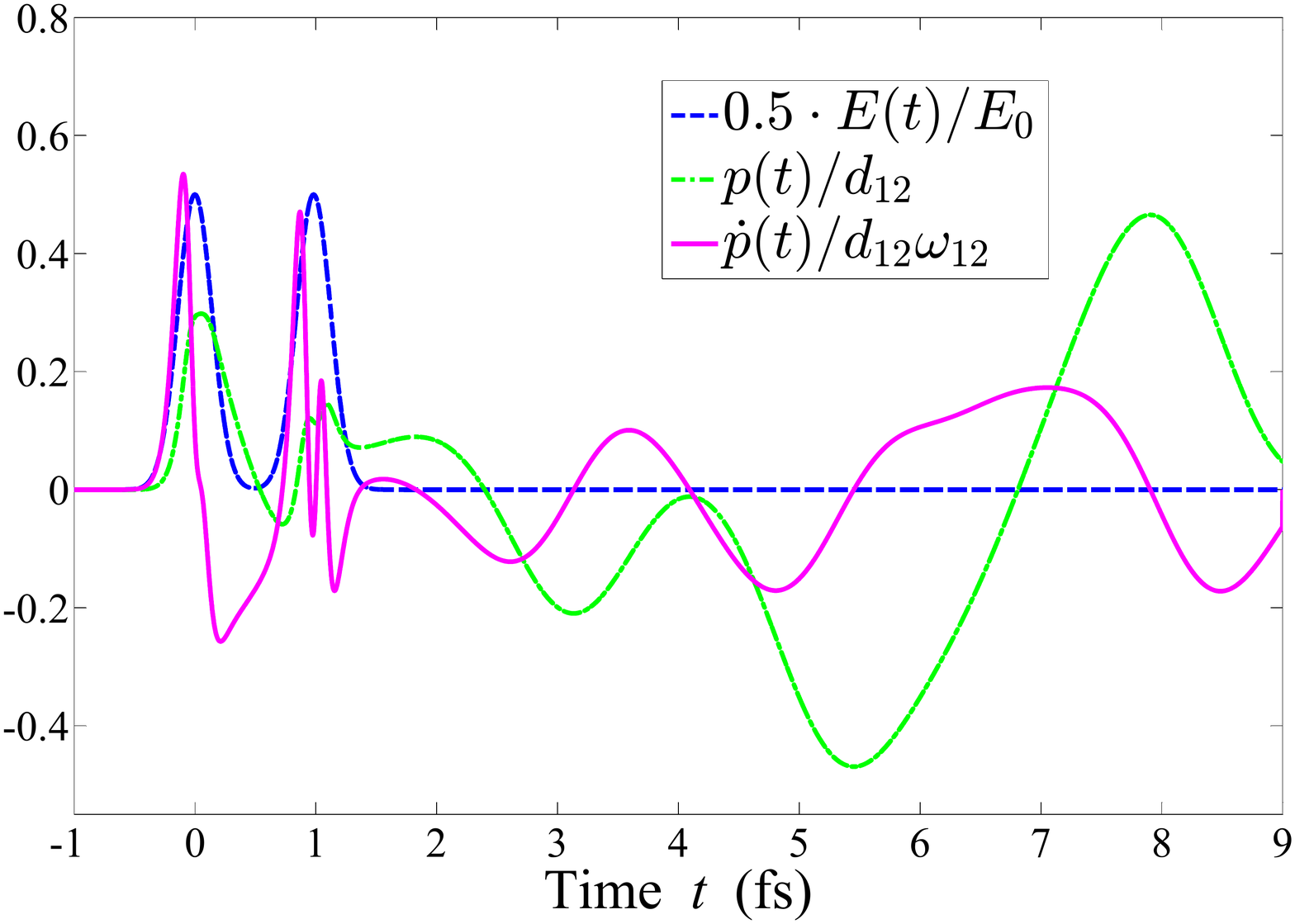}
\caption{(Color online) The induced polarization per a single atom (dashed green curve) together with its temporal derivative (magenta solid curve) and the electric field of the excitation pulses (black solid line); the pulse amplitude $E_0 = 10^6$ ESU, all other parameters are the same as in Fig.~\ref{fig2}. All quantities are plotted in dimensionless units.}
\label{fig8}
\end{figure}
%%%%%%%%%%%%%%%%%%%%%%%%%%%%%%%%%%%%%%%%%%%%%%%%%%%%%%

%%%%%%%%%%%%%%%%%%%%%%%%%%%%%%%%%%%%%%%%%%%%%%%%%%%%%
\begin{figure}[tpb]
\centering
\includegraphics[width=1.0\linewidth]{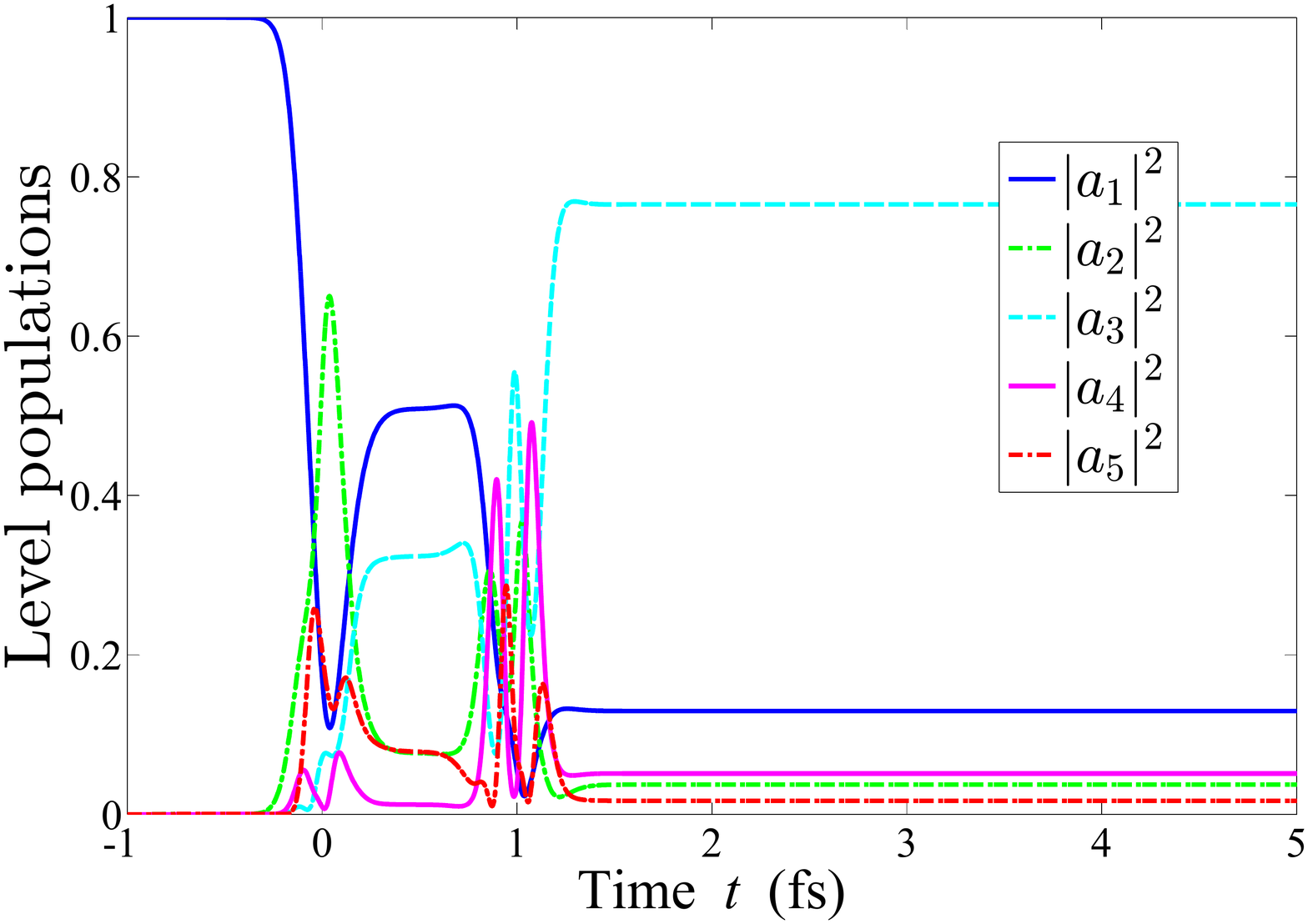}
\caption{(Color online) The populations of atom's levels vs. time; all  parameters are the same as in Fig.~\ref{fig8}. }
\label{fig9}
\end{figure}
%%%%%%%%%%%%%%%%%%%%%%%%%%%%%%%%%%%%%%%%%%%%%%%%%%%%%%

The obtained value Eq.~\eqref{threshold} thus provides the threshold value of the amplitude of the driving pulses in Eq.~\eqref{Et}. For the field strength well below this threshold the two-level approximation can be reliably used and the effect of higher excited levels can be safely neglected. On the other hand, for the pulse amplitude well above the threshold the contributions of higher excited levels play important role in the medium response and have to be taken into account. It is interesting to note that the product $E_{\rm{thr}} \tau_p$ in Eq.~\eqref{criterion} matches the respective pulse area Eq.~\eqref{area} up to a constant factor, while the ratio $\hbar / d_{1k}$ represents the so-called atomic sale of the electric pulse area introduced in Ref.~\cite{Arkhipov_JL_2021}. This atomic scale refers to the characteristic value of the electric pulse area needed to excite an atomic system by a driving unipolar pulse. The criterion Eq.~\eqref{criterion} can be therefore interpreted in such a way, that the two-level description stays valid until the electric area of the driving pulses becomes comparable to its atomic scale for the respective multi-level atom.

The estimated value in Eq.~\eqref{threshold} appears to coincide with the pulse amplitude in Fig.~\ref{fig6}. The case shown in  Figs.~\ref{fig6}-\ref{fig7} therefore refers to the intermediate case in terms of the applicability of the two-level treatment, what is indeed in good agreement with the behaviour seen in Figs.~\ref{fig6}-\ref{fig7}.

Further increasing the amplitude of the driving pulses Eq.~\eqref{Et} beyond the threshold value Eq.~\eqref{threshold} leads to the even larger contribution of other polarization components beyond the main one $p_{12}$ and respectively stronger residual field oscillations. As the result, the emission of the atom has already nothing to do with a single-cycle pulse shown in Fig.~\ref{fig2}, but rather takes the form of a complex non-structured field oscillations without any characteristic subcycle parts.

%%%%%%%%%%%%%%%%%%%%%%%%%%%%%%%%%%%%%%%%%%%%%%%%%%%%%
\begin{figure}[tpb]
\centering
\includegraphics[width=1.0\linewidth]{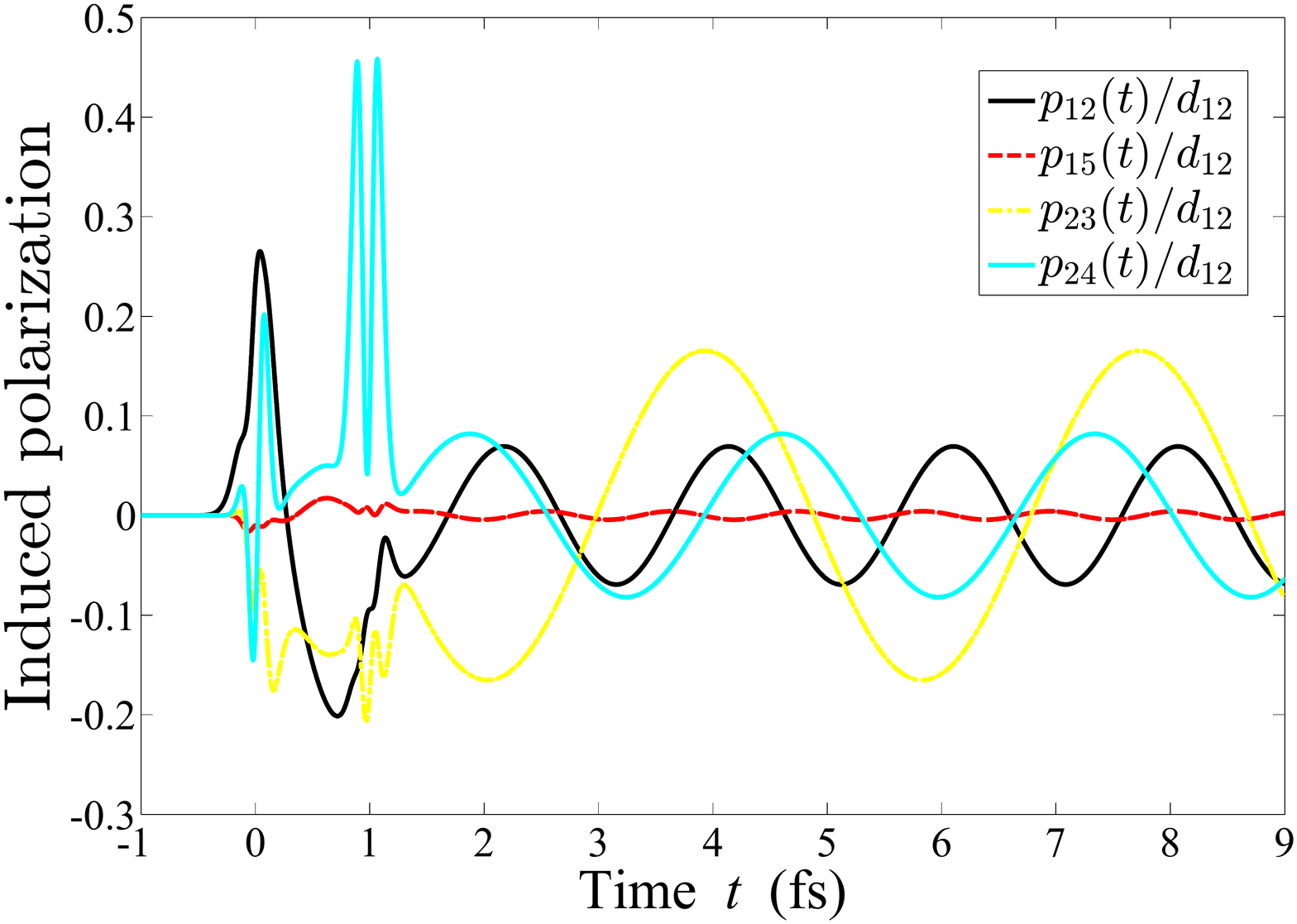}
\caption{(Color online) The strongest polarization components of a single atom vs. time; all  parameters are the same as in Fig.~\ref{fig8}. }
\label{fig10}
\end{figure}
%%%%%%%%%%%%%%%%%%%%%%%%%%%%%%%%%%%%%%%%%%%%%%%%%%%%%%

An example of such emission is shown in Fig.~\ref{fig8}, where the pulse amplitude was increased to $E_0 = 10^6$ ESU as compared to Fig.~\ref{fig6}. The respective dynamics of the level populations in Fig.~\ref{fig9} exhibits now strong depletion of the ground state after the action of both pulses Eq.~\eqref{Et}. The high residual populations of excited levels are also to be noted, which result from the complex dynamics of transitions between levels in so intense fields. In particularly, the level 3 (4S) becomes the most populated after the passage of driving pulses with the residual population several times larger than in the ground state. The separate contributions of strongest polarization components are plotted in Fig.~\ref{fig10}. First of all, one can see that the polarization component of the main transition $p_{12}$ indeed ceases to be the strongest one and other components, such as $p_{23}$ and $p_{24}$, start to dominate in the medium response. Besides, the strong interplay of all excited levels leads to large residual oscillating tails even at the main component $p_{12}$. This fact means that the second subcycle pulse in Eq.~\eqref{Et} can not be anymore used to stop the induced medium oscillations, so that the coherent control of multi-level atoms turns out unfeasible.

In the case of such an intense driving field the atom emission can no longer be even qualitatively described using the two-level approximation. Hence, one can conclude that the optical response of a multi-level medium driven by subcycle pulses, including unipolar ones, can be well modelled within the two-level model, but just for excitation amplitudes much less a certain threshold. As long as such inequality does not hold anymore, the higher energy levels have to be taken into account as well.

%%%%%%%%%%%%%%%%%%%%%%%%%%%%%%%%%%%%%%%%%%%%%%%%%%%%%
\begin{figure}[tpb]
\centering
\includegraphics[width=1.0\linewidth]{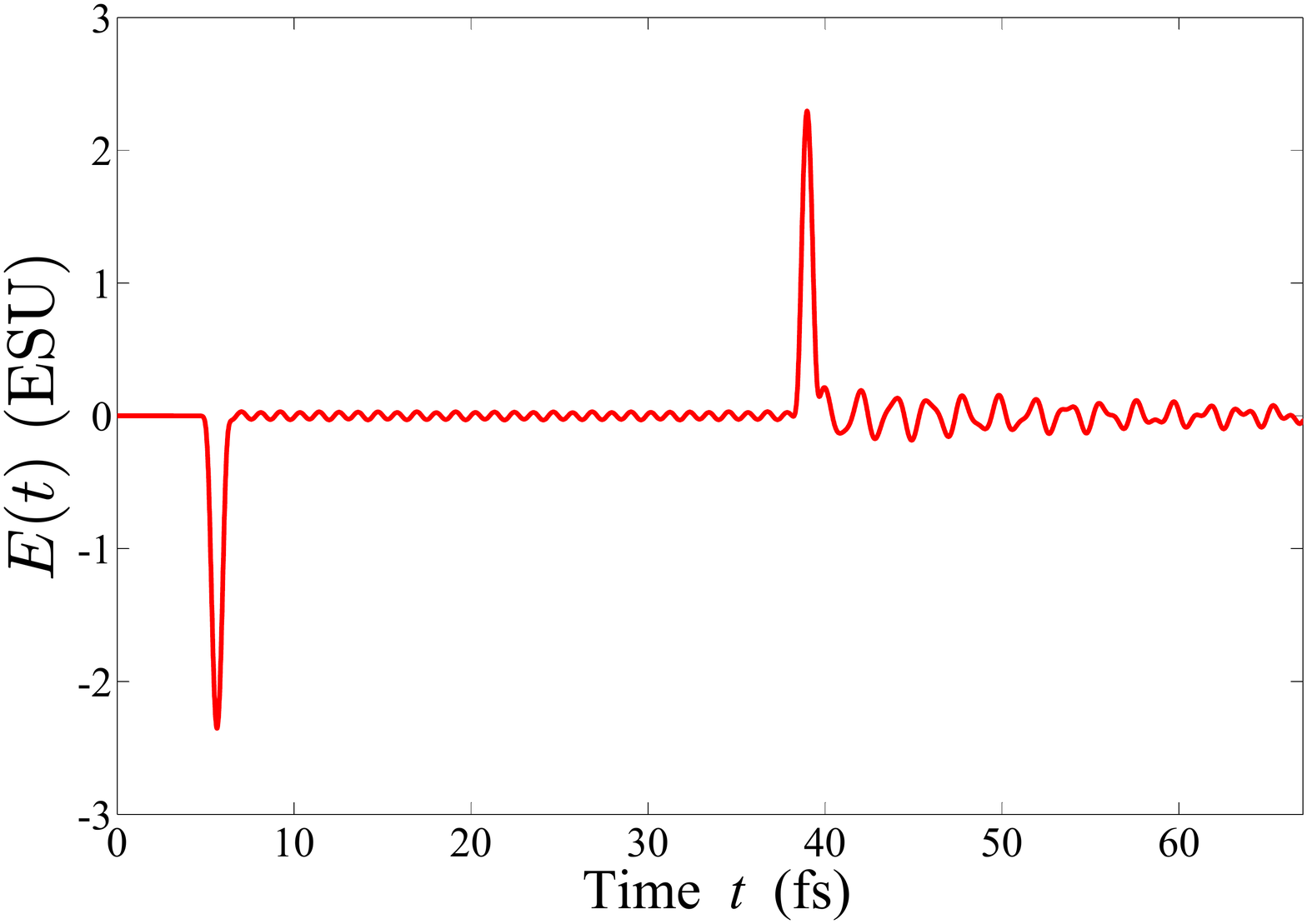}
\caption{(Color online) Reflected electric field from an extended medium layer; the density of the resonant atoms $N = 3 \cdot 10^{18}$ cm$^{-3}$, the layer thickness $L = 5 \ \mu$m, the excitation pulses have the amplitude $E_0 = 10^4$ ESU and the duration $\tau_p = 200$ as.}
\label{fig11}
\end{figure}
%%%%%%%%%%%%%%%%%%%%%%%%%%%%%%%%%%%%%%%%%%%%%%%%%%%%%%

These findings, however, can appear to confront the common thinking that the two-level model is only valid, when a long multi-cycle exciting pulse with near-resonant carrier frequency is used. It is therefore worth noting that the treatment above was based on the assumption of the duration of driving pulses less than the medium's transition periods. It is thus only this limiting case, when the coherent control of the multi-level medium response can be efficiently implemented and the abovesaid validity of the two-level approximation can be justified. In the opposite limiting case of relatively long driving pulses the medium response can be indeed described by the two-level model just as long as the pulse carrier frequency coincides with the frequency of the respective transition.

\section{Tailoring the response of a thick medium layer}

Let us now consider the response of an extended layer of a multi-level resonant medium. Namely, we assume a layer thickness to largely exceed the wavelengths of the resonant transitions in the medium. In this case we have to solve the equations Eq.~\eqref{Bloch} coupled to the wave equation Eq.~\eqref{eq_wave} through the expression Eq.~\eqref{polariz}. The wave equation Eq.~\eqref{eq_wave} was solved using the finite-difference time-domain method (FDTD), while for the equations Eq.~\eqref{Bloch} we applied the 4-th order Runge-Kutta scheme. The rectangular numerical grid with the spatial grid spacing $\Delta x = 3$ nm and the temporal spacing $\Delta t = \Delta x / c = 0.01$ fs was used. We imposed transparent boundary conditions for the electric field at the boundaries of the computational domain. The  field emitted by the medium layer in reflection was recorded at the left boundary of the computational area 1 $\mu$m away from the left edge of the medium layer.

%%%%%%%%%%%%%%%%%%%%%%%%%%%%%%%%%%%%%%%%%%%%%%%%%%%%%
\begin{figure}[tpb]
\centering
\includegraphics[width=1.0\linewidth]{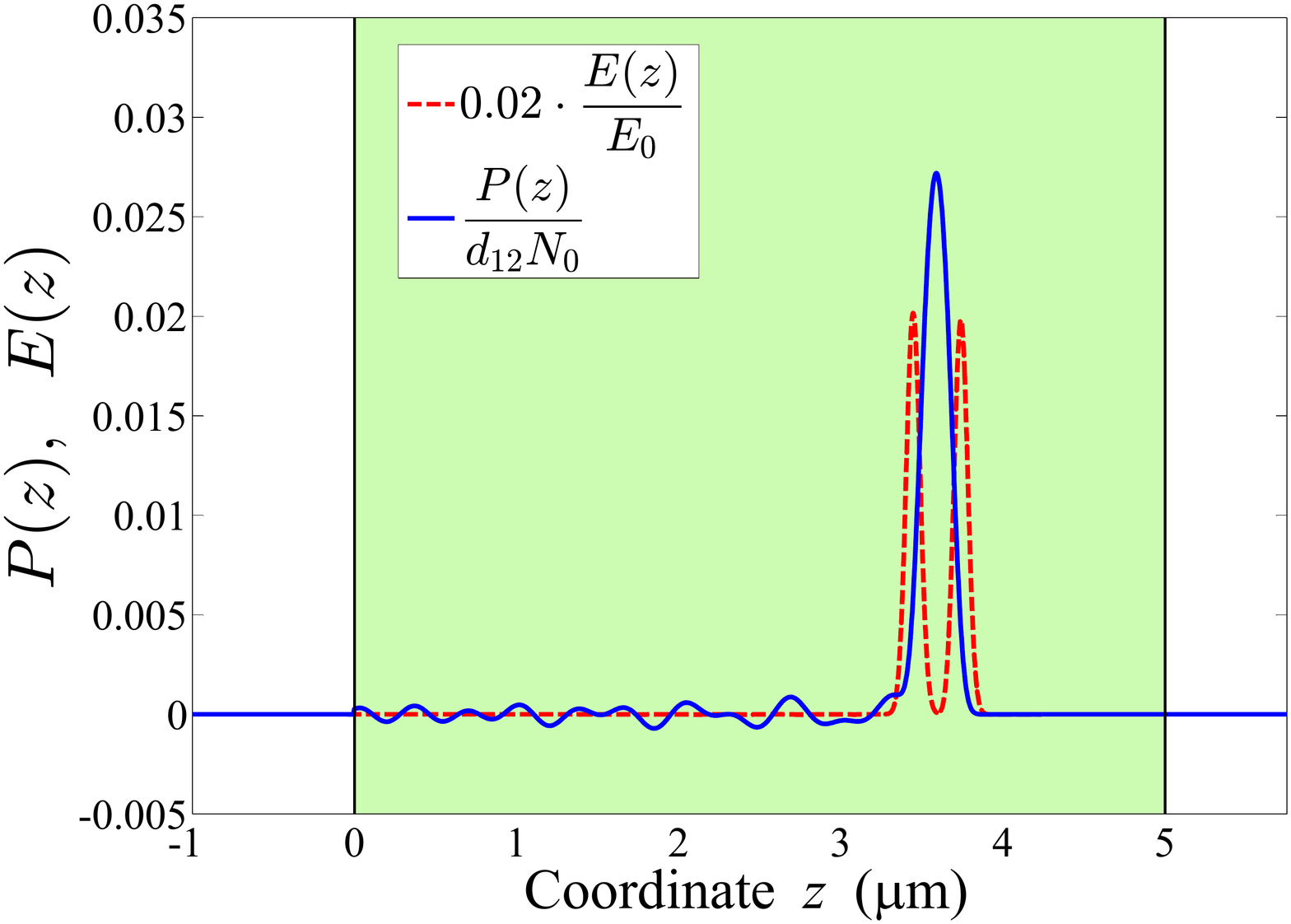}
\caption{(Color online) The instantaneous spatial distribution of the induced medium polarization inside an extended layer of the multi-level medium of the thickness $L = 5 \ \mu$m (blue solid line) for the case of Fig.~\ref{fig11}, together with the distribution of the electric field (red dashed line). The plotted distributions correspond to a time point when both excitation pulses Eq.~\eqref{Et} are  propagating inside the medium layer. The shaded area indicates the medium layer. All parameters are the same as in Fig.~\ref{fig11}. Both the electric field and the medium polarization are properly rescaled and plotted in dimensionless units.}
\label{fig12}
\end{figure}
%%%%%%%%%%%%%%%%%%%%%%%%%%%%%%%%%%%%%%%%%%%%%%%%%%%%%%

Fig.~\ref{fig11} shows the emitted field in reflection from a low-pressure layer of sodium atoms of the thickness 5 $\mu$m, when excited by two unipolar pulses Eq.~\eqref{Et}. The pulse amplitude equals $E_0 = 10^4$ ESU, what corresponds to the response of a single atom shown in Fig.~\ref{fig2}. We deliberately pick the the density of the resonant atoms $N$ to be low enough, so that the excitation pulses Eq.~\eqref{Et} experience negligible transformation upon their propagation through the layer. Otherwise the second excitation pulse would not be able to stop the induced polarization oscillations like in Fig.~\ref{fig2} and we would end up with strong and irregular residual field oscillations in the emitted field.

The obtained field in Fig.~\ref{fig11} yields two unipolar pulses of opposite polarity and the duration $\pi / \omega_{12}$ separated by the long plateau with the near-zero electric field. Such sort of emission arises due to the interference of the single-cycle pulses like in  Fig.~\ref{fig2} produced by different slices along the layer. As the result, such single-cycle pulses fully compensate each other leading to the near-zero field except for the very leading and the very trailing edges of the layer response, where the constructive interference of the half-cycles of the same polarity takes place \cite{Pakhomov_PRA_2022}. The time delay between both unipolar bursts is proportional to the thickness of the medium layer. Note that a similar response was recently obtained from an extended layer of a simplest two-level medium excited by a pair of unipolar pulses in Ref.~\cite{Pakhomov_JETPL}.

%%%%%%%%%%%%%%%%%%%%%%%%%%%%%%%%%%%%%%%%%%%%%%%%%%%%%
\begin{figure}[tpb]
\centering
\includegraphics[width=1.0\linewidth]{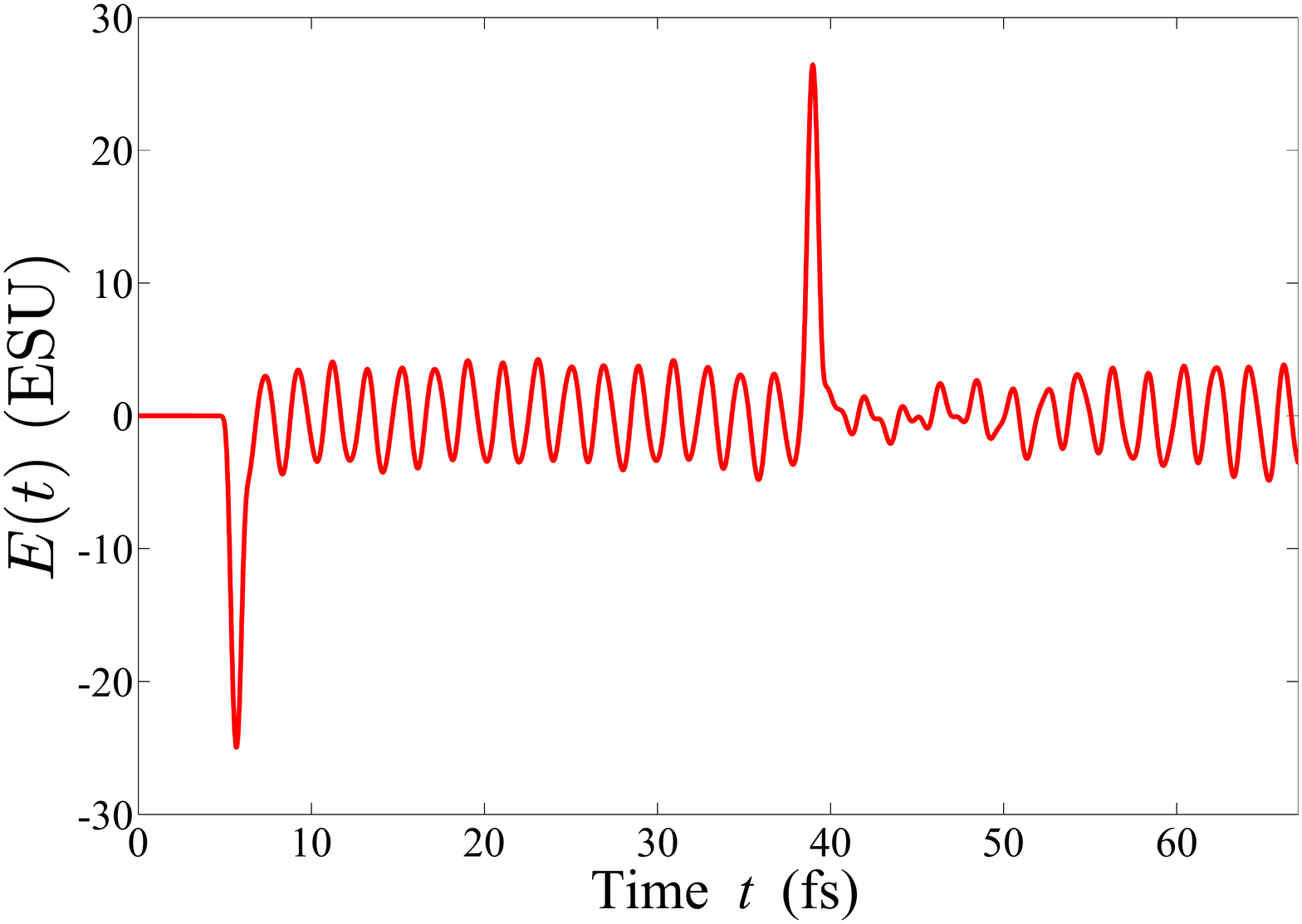}
\caption{(Color online) Reflected electric field from an extended medium layer; the excitation pulses have the amplitude $E_0 = 10^5$ ESU, all other parameters are the same as in Fig.~\ref{fig11}.}
\label{fig13}
\end{figure}
%%%%%%%%%%%%%%%%%%%%%%%%%%%%%%%%%%%%%%%%%%%%%%%%%%%%%%

Fig.~\ref{fig12} illustrates the respective spatial distribution of the induced medium polarization inside the layer for the parameters from Fig.~\ref{fig11}. One can see in Fig.~\ref{fig12} exactly the same half-cycle burst of the medium polarization, as the one in Fig.~\ref{fig2} for a single atom. In the case of an extended medium layer, however, this polarization burst propagates along the medium layer, while being "sandwiched" in between two half-cycle driving pulses Eq.~\eqref{Et}. The underlying dynamics of the medium response is completely identical to those shown in Fig.~\ref{fig2}, i.e. the first driving pulse induces the medium polarization oscillations mainly at the frequency $\omega_{21}$, and the second driving pulse almost fully stops these oscillations, resulting in an isolated polarization burst. The half-cycle burst of the medium polarization in Fig.~\ref{fig12} arises over the time slot, when the first one of the driving pulses Eq.~\eqref{Et} has entered the medium but the second one has not yet. In this time the first unipolar half-cycle pulse in Fig.~\ref{fig11} gets emitted. Then, as both half-cycle driving pulses propagate inside the layer with the polarization burst in between them like shown in Fig.~\ref{fig12}, almost no emission appears (this corresponds to the interval in between both half-cycle pulses in Fig.~\ref{fig11}). Finally, while
the first driving pulse has exited the medium but the second one is still inside, the half-cycle burst of the medium polarization in Fig.~\ref{fig12} gradually vanishes, what results in the emission of the second unipolar half-cycle pulse in Fig.~\ref{fig11}.

%%%%%%%%%%%%%%%%%%%%%%%%%%%%%%%%%%%%%%%%%%%%%%%%%%%%%
\begin{figure}[tpb]
\centering
\includegraphics[width=1.0\linewidth]{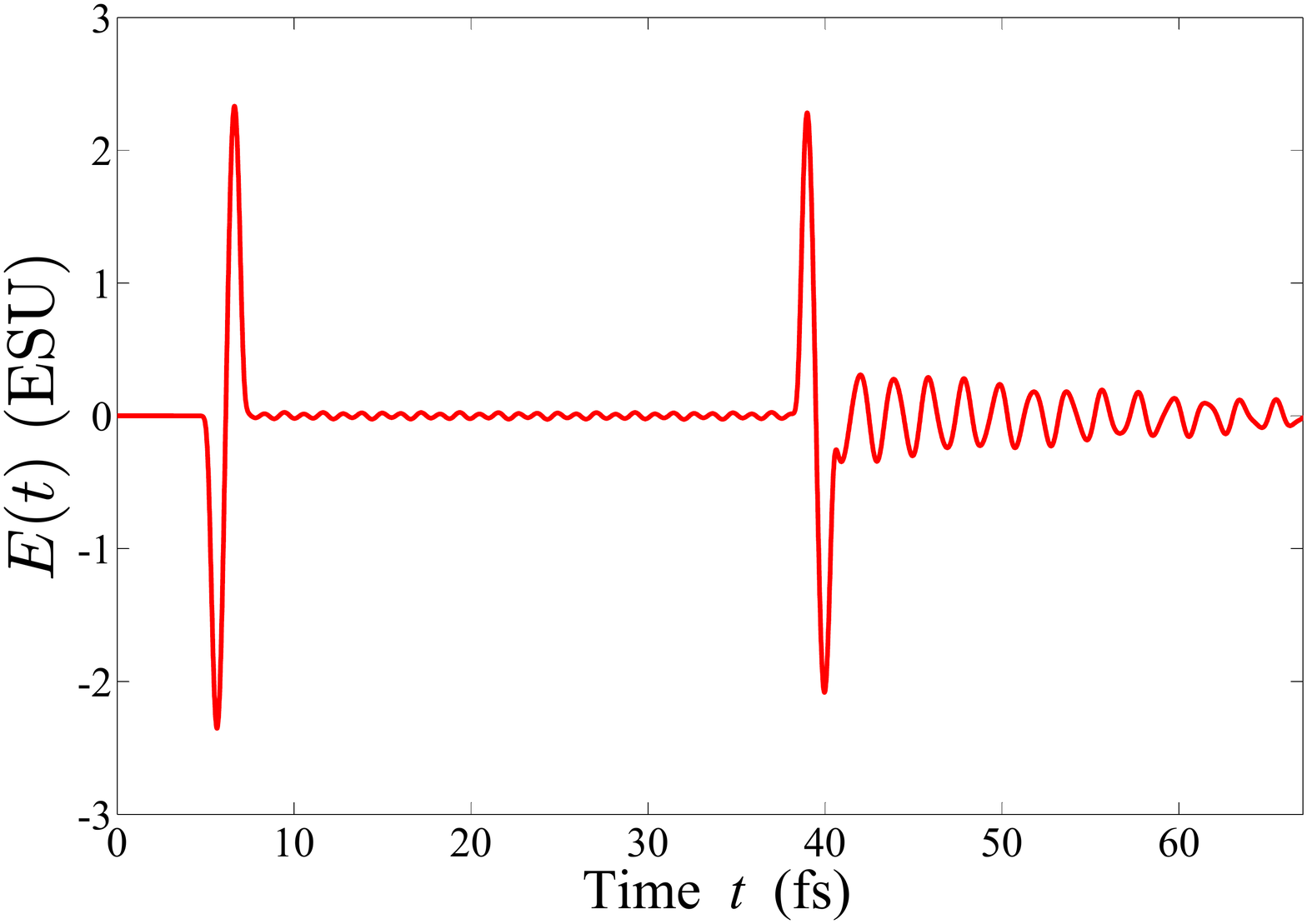}
\caption{(Color online) Reflected electric field from an extended medium layer; the parameters are the same as in Fig.~\ref{fig11}, but the exciting unipolar pulses are taken of different polarity.}
\label{fig14}
\end{figure}
%%%%%%%%%%%%%%%%%%%%%%%%%%%%%%%%%%%%%%%%%%%%%%%%%%%%%%

With increasing the pulse peak amplitude the residual oscillations become stronger and distort the above described medium response. An example plot in Fig.~\ref{fig13} gives the layer emission when the peak amplitude of the driving pulses was $E_0 = 10^5$ ESU. The respective response of a single atom is demonstrated in Fig.~\ref{fig6}. The unipolar bursts are still well pronounced in Fig.~\ref{fig13}, but the field oscillations both in between the unipolar half-sines and after the second unipolar burst are now much more intense and comparable in amplitude to the main unipolar peaks.

Fig.~\ref{fig14} illustrates another case, when the excitation pulses have opposite polarity, i.e. instead of Eq.~\eqref{Et} we take:
\begin{eqnarray}
\nonumber
E(t) &=&  E_0 e^{-t^2/\tau_p^2} - E_0 e^{-(t - \Delta)^2/\tau_p^2}.
\label{Et_2}
\end{eqnarray}
In this case, however, the value of the pulse-to-pulse delay provided by Eq.~\eqref{delta} is no more applicable, since the second pulse of opposite polarity would now just speed up the polarization oscillations instead of stopping them. The time delay between pulses $\Delta$ must be rather taken as:
\begin{eqnarray}
\nonumber
\Delta = T_{12} = \frac{2 \pi}{\omega_{12}},
\label{delta_2}
\end{eqnarray}
i.e. the full period of the resonant medium oscillations at the frequency of the main transition. As one can see in Fig.~\ref{fig14} the field interference leads to a pair of single-cycle pulses in the layer emission, which are separated again by the time interval with almost no field.

%%%%%%%%%%%%%%%%%%%%%%%%%%%%%%%%%%%%%%%%%%%%%%%%%%%%%
\begin{figure}[tpb]
\centering
\includegraphics[width=1.0\linewidth]{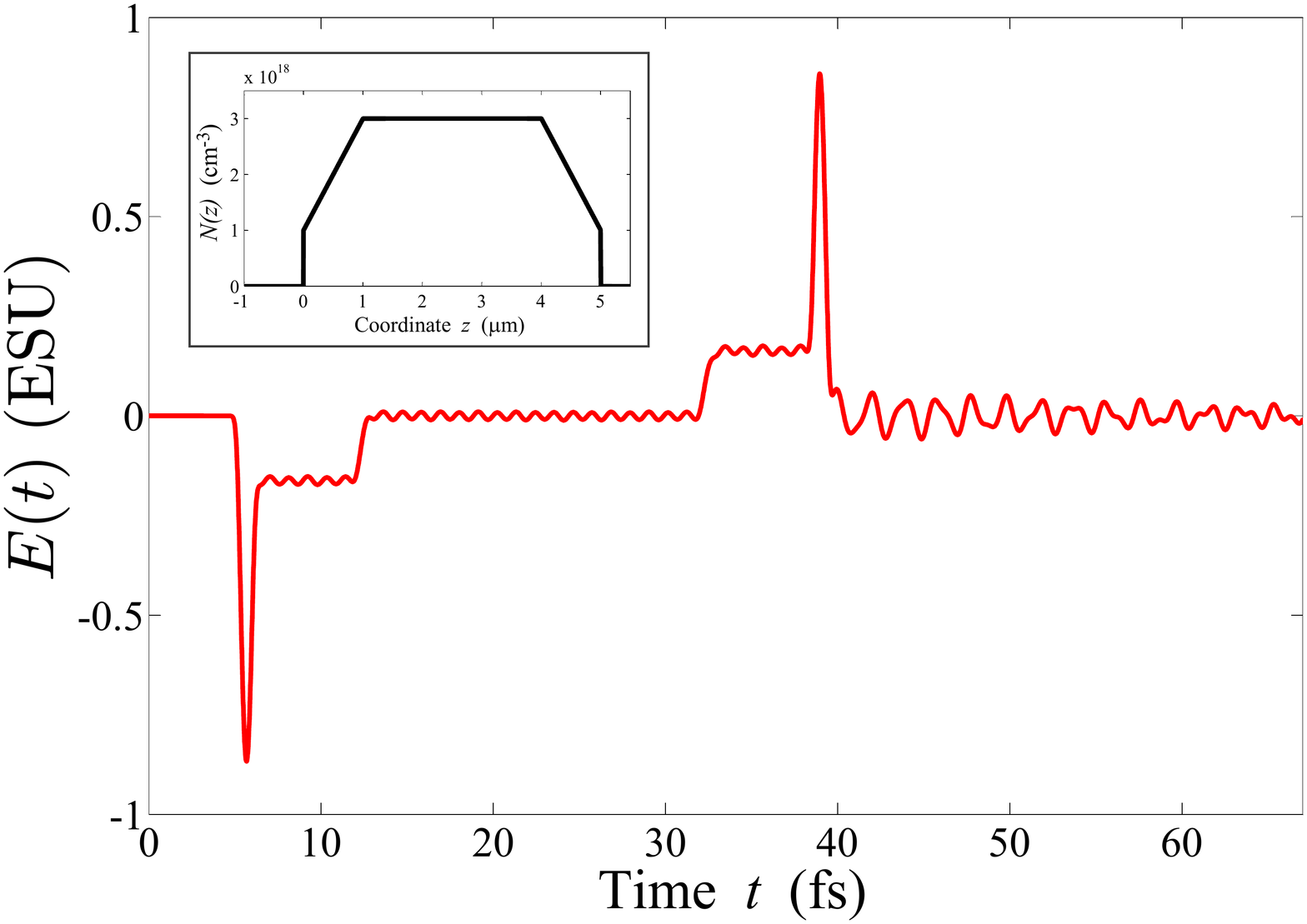}
\caption{(Color online) Reflected electric field from an extended medium layer with the non-homogeneous trapezoidal-shaped spatial density profile; the density of the resonant atoms $N(z)$ linearly grows from $1 \cdot 10^{18}$ cm$^{-3}$ till $3 \cdot 10^{18}$ cm$^{-3}$ across the left-side part of the layer of the thickness 1 $\mu$m, stays constant and equal to $3 \cdot 10^{18}$ cm$^{-3}$ in the central part of the layer of the thickness 3 $\mu$m, and then linearly decreases from $3 \cdot 10^{18}$ cm$^{-3}$ till $1 \cdot 10^{18}$ cm$^{-3}$
across the right-side part of the layer of the thickness 1 $\mu$m (shown in the inset); other parameters are the same as in Fig.~\ref{fig11}.}
\label{fig15}
\end{figure}
%%%%%%%%%%%%%%%%%%%%%%%%%%%%%%%%%%%%%%%%%%%%%%%%%%%%%%

%%%%%%%%%%%%%%%%%%%%%%%%%%%%%%%%%%%%%%%%%%%%%%%%%%%%%
\begin{figure}[tpb]
\centering
\includegraphics[width=1.0\linewidth]{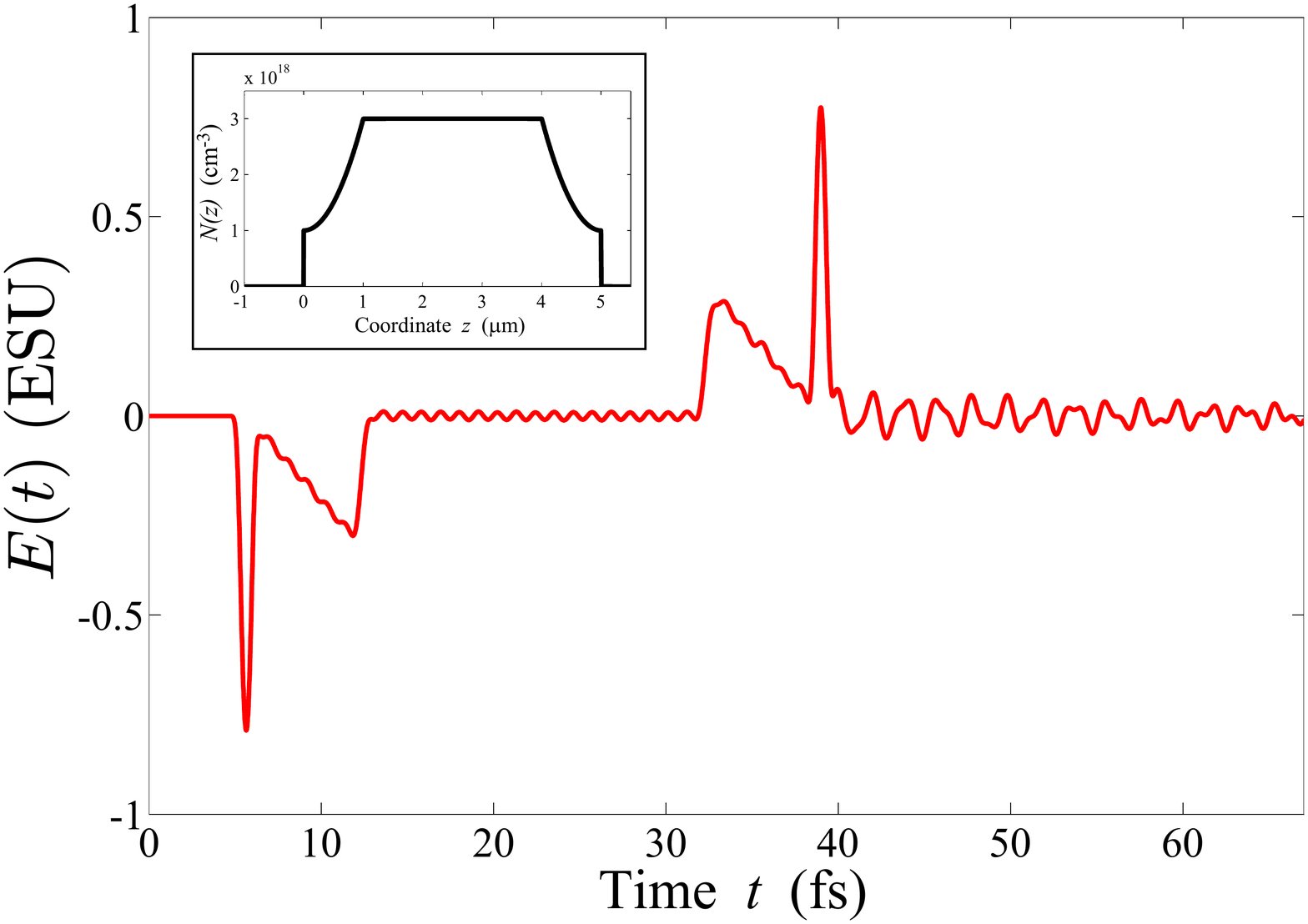}
\caption{(Color online) Reflected electric field from an extended medium layer with the non-homogeneous parabolic flat-top spatial density profile; the density of the resonant atoms $N(z)$ quadratically grows from $1 \cdot 10^{18}$ cm$^{-3}$ till $3 \cdot 10^{18}$ cm$^{-3}$ across the left-side part of the layer of the thickness 1 $\mu$m, stays constant and equal to $3 \cdot 10^{18}$ cm$^{-3}$ in the central part of the layer of the thickness 3 $\mu$m, and then quadratically decreases from $3 \cdot 10^{18}$ cm$^{-3}$ till $1 \cdot 10^{18}$ cm$^{-3}$
across the right-side part of the layer of the thickness 1 $\mu$m (shown in the inset); other parameters are the same as in Fig.~\ref{fig11}.}
\label{fig16}
\end{figure}
%%%%%%%%%%%%%%%%%%%%%%%%%%%%%%%%%%%%%%%%%%%%%%%%%%%%%%

Next, following the findings of Ref.~\cite{Pakhomov_PRA_2022} for Raman-active media, we are interested to consider the emitted field from an extended layer with a spatially inhomogeneous density distribution along the layer thickness. This could be achieved, e.g. by cooling or heating of a gas cell from one side. We begin with a linearly varying spatial density profile. Specifically, we take a trapezoidal-shaped density, which linearly grows near the left side of the layer, stays constant at the central part and then linearly decreases again near the right side of the layer. The respective simulation results are depicted in Fig.~\ref{fig15}. Here one see besides the unipolar bursts also rectangular-shaped unipolar pulses placed right next to the stronger half-sine peaks. While the half-wave bursts arise due to the discontinuity of the spatial density at the edges of the layer, the rectangular unipolar parts arise thanks to the interference of the emitted fields like in Fig.~\ref{fig2} within the linearly varying density profile.

%%%%%%%%%%%%%%%%%%%%%%%%%%%%%%%%%%%%%%%%%%%%%%%%%%%%%
\begin{figure}[tpb]
\centering
\includegraphics[width=1.0\linewidth]{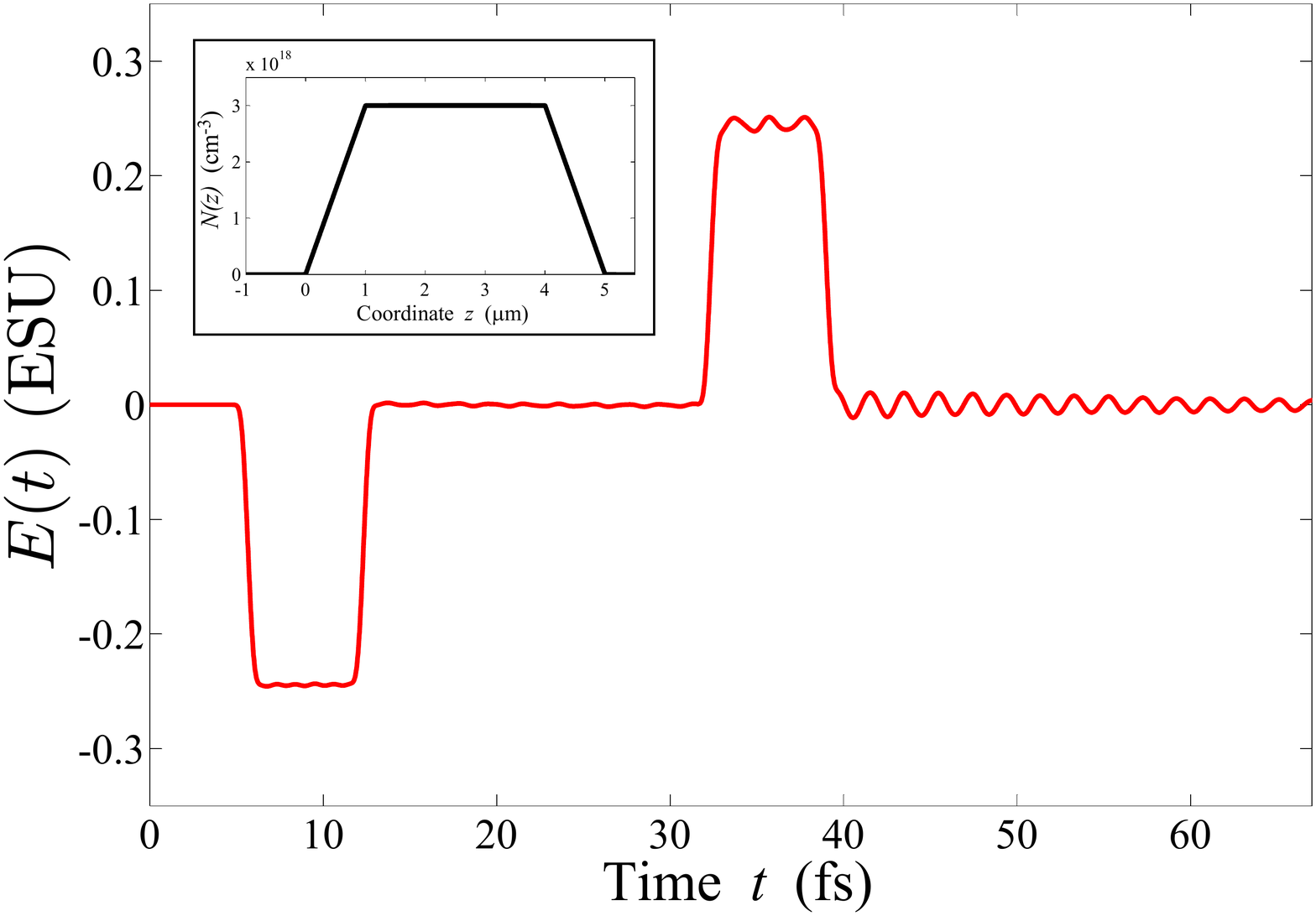}
\caption{(Color online) Reflected electric field from an extended medium layer with the non-homogeneous trapezoidal-shaped spatial density profile; the density of the resonant atoms $N(z)$ linearly grows from $0$ till $3 \cdot 10^{18}$ cm$^{-3}$ across the left-side part of the layer of the thickness 1 $\mu$m, stays constant and equal to $3 \cdot 10^{18}$ cm$^{-3}$ in the central part of the layer of the thickness 3 $\mu$m, and then linearly decreases from $3 \cdot 10^{18}$ cm$^{-3}$ till $0$ across the right-side part of the layer of the thickness 1 $\mu$m (shown in the inset); other parameters are the same as in Fig.~\ref{fig11}.}
\label{fig17}
\end{figure}
%%%%%%%%%%%%%%%%%%%%%%%%%%%%%%%%%%%%%%%%%%%%%%%%%%%%%%

The exact shape of the medium response can be analytically calculated using the approach from Ref.~\cite{Pakhomov_PRA_2022}, where unipolar pulses of various waveforms were obtained from an optically thick layer of a Raman-active medium excited by a pair of multi-cycle bipolar pulses. The basic idea is to approximate the emitted field from a thin medium slice in Fig.~\ref{fig2} as a single period of the harmonic function of the frequency $\omega_{21}$ and to sum up such emitted waves across the whole layer. We then end up with the expression for the layer emission of the form: 
\begin{eqnarray}
\nonumber
E(t) &\sim& \int_0^L N(z) \ \sin \omega_{21} \Big( t - \frac{2z + D}{c} \Big) \cdot \\
\nonumber
&& \Big[ \Theta \Big( t - \frac{2z + D}{c} \Big)  -  
\Theta \Big( t - \frac{2z + D}{c} - T_{12} \Big) \Big] dz, \\
\label{Emission}
\end{eqnarray}
with the spatial density $N(z)$, the Heaviside step function $\Theta$ and the distance $D$ from the layer to the detector. According to the findings in Ref.~\cite{Pakhomov_PRA_2022}, the power-law function $N(z) \sim z^n$ leads to the emitted field in Eq.~\eqref{Emission} with the  power-law temporal dependence $E(t) \sim t^{n-1}$. That is why the linearly increasing or decreasing spatial density results in the rectangular unipolar emitted field, like in Fig.~\ref{fig15}.

To further illustrate this finding, we move next to another spatial density profile and take a flat-top profile with the parabolic dependence at the edges instead of the linear one in Fig.~\ref{fig15}. The simulation result for this profile is plotted in Fig.~\ref{fig16}. One can see two triangular-shaped unipolar pulses of opposite polarity in the layer emission. The time delay between both triangular parts is again like in Fig.~\ref{fig11} proportional to the thickness of the medium layer. Next to the triangular pulses we get also the half-cycle unipolar peaks because of the jumps in the spatial density distribution at the layer edges. These half-cycle unipolar bursts can be fully eliminated, when assuming the spatial density $N(z)$ to be equal zero at the layer edges and quadratically growing towards the layer center.

\begin{figure}[tb]
\centering
\includegraphics[width=1.0\linewidth]{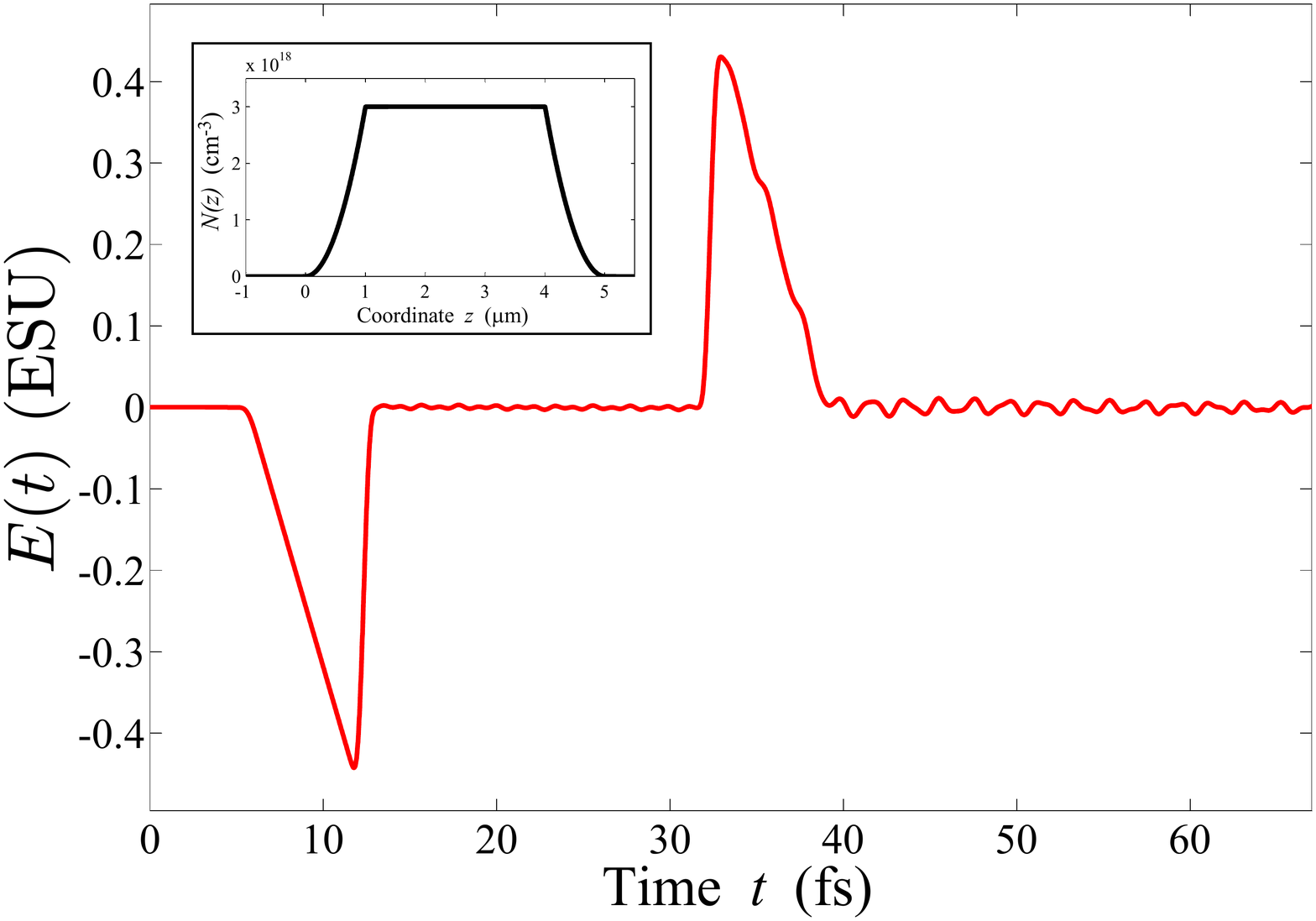}
\caption{(Color online) Reflected electric field from an extended medium layer with the non-homogeneous parabolic flat-top spatial density profile; the density of the resonant atoms $N(z)$ quadratically grows from $0$ till $3 \cdot 10^{18}$ cm$^{-3}$ across the left-side part of the layer of the thickness 1 $\mu$m, stays constant and equal to $3 \cdot 10^{18}$ cm$^{-3}$ in the central part of the layer of the thickness 3 $\mu$m, and then quadratically decreases from $3 \cdot 10^{18}$ cm$^{-3}$ till $0$ across the right-side part of the layer of the thickness 1 $\mu$m (shown in the inset); other parameters are the same as in Fig.~\ref{fig11}.}
\label{fig18}
\end{figure}

Similar case is shown in Fig.~\ref{fig17}, where we take again the flat-top trapezoidal density profile as in Fig.~\ref{fig15}, but now assume the spatial density $N(z)$ to be equal zero at the layer edges and linearly growing towards the central section of the layer. Such type of the density profile with the density dropping to zero at the edges could be created, e.g. upon the flow of a gas through a finite-size nozzle. As can be seen from Fig.~\ref{fig17}, we indeed get rid of the half-cycle unipolar bursts at the edges of the medium emission, since the medium density now changes continuously across the layer borders. Thereby we end up with almost exactly rectangular unipolar pulses of several fs in duration. Finally, Fig.~\ref{fig18} corresponds to the flat-top profile with the parabolic dependence near the edges as in Fig.~\ref{fig16}, but with the density dropping exactly to zero at the edges. Here one can see a pair of triangular unipolar optical pulses of opposite polarity, similar to the emitted field in Fig.~\ref{fig16} but without half-cycle bursts.

\begin{figure}[tb]
\centering
\includegraphics[width=1.0\linewidth]{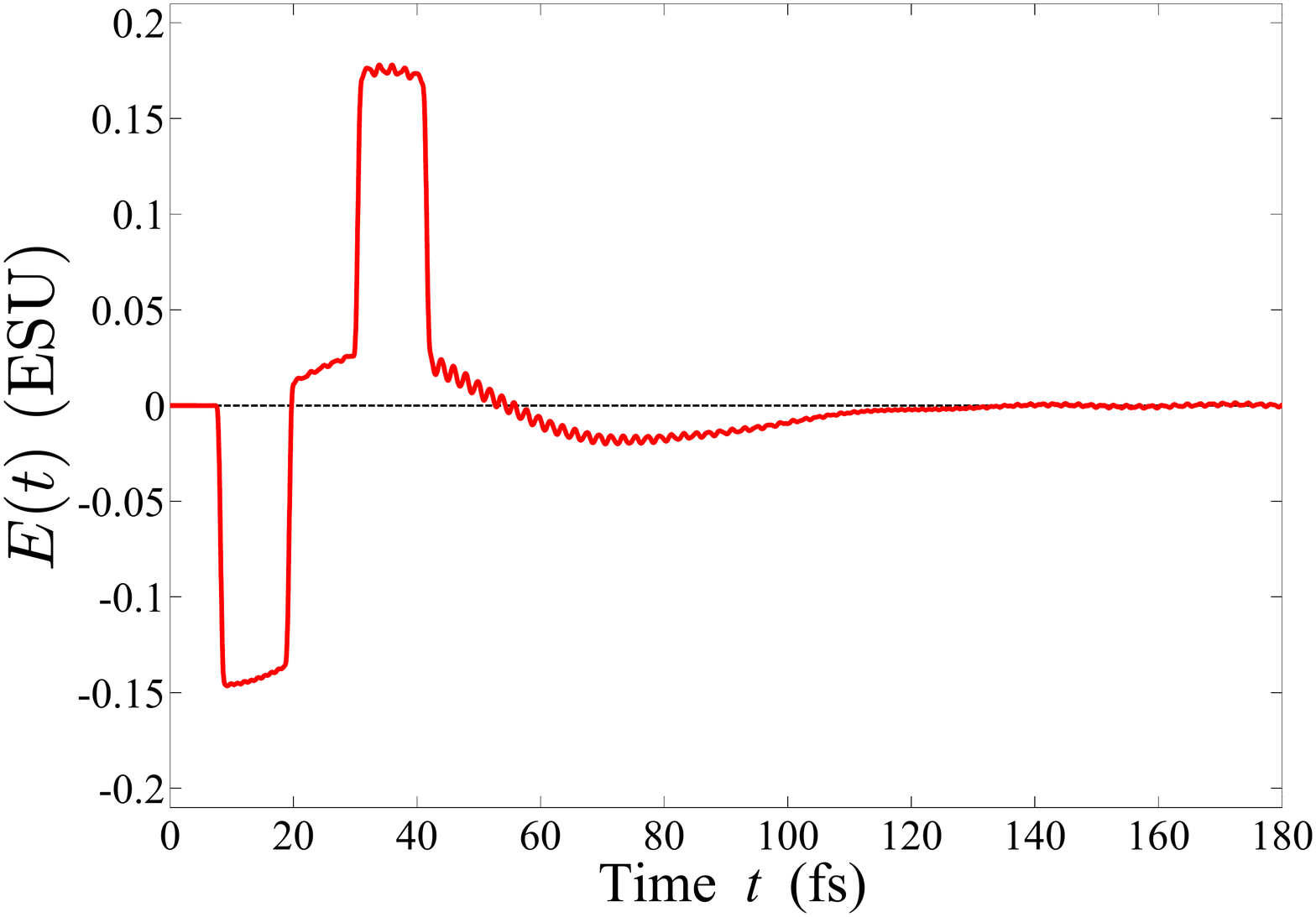}
\caption{(Color online) Reflected electric field from an extended medium layer with the same trapezoidal-shaped spatial density profile as in Fig.~\ref{fig17}, but excited by a pair of quasi-unipolar half-cycle pulses Eq.~\eqref{Et_1}; the  parameters of the driving pulses are the same as in Fig.~\ref{fig3-2}.} 
\label{fig19}
\end{figure}

Lastly, we would like to show that the obtained  above waveforms can be also produced, when using more feasible quasi-half-cycle driving pulses Eq.~\eqref{Et_1} as in Fig.~\ref{fig3-2}. An example of the calculated medium emission is plotted in Fig.~\ref{fig19} for the same trapezoidal-shaped spatial density profile as in Fig.~\ref{fig17} and the same quasi-unipolar driving pulses as in Fig.~\ref{fig3-2}. One can see again a pair of rectangular-shaped unipolar pulses similar to those in Fig.~\ref{fig17}, although some discrepancies from the perfect rectangular profiles are visible caused by the presence of the long tails of the exciting pulses.

It should be emphasized that the medium emission in Figs.~\ref{fig11},~\ref{fig15}-\ref{fig19} consists of 2 unipolar parts of identical shape but opposite polarity. Thus, the total electric pulse area Eq.~\eqref{area} of the emitted field equals zero. This means that the medium emission does not comprise the zero-frequency spectral component. At the same time these unipolar parts can be well separated in time in order to be used in possible applications.

%%%%%%%%%%%%%%%%%%%%%%%%%%%%%%%%%%%%%%%%%%%%%%%%%%%%%
%%%%%%%%%%%%%%%%%%%%%%%%%%%%%%%%%%%%%%%%%%%%%%%%%%%%%%

%%%%%%%%%%%%%%%%%%%%%%%%%%%%%%%%%%%%%%%%%%%%%%%%%%%%%
%%%%%%%%%%%%%%%%%%%%%%%%%%%%%%%%%%%%%%%%%%%%%%%%%%%%%%

In general, more complex waveshapes of the emitted field can be produced by proper adjusting the density profile $N(z)$. Similar to the findings of the previous section, this result only holds for the driving pulse amplitudes below a certain threshold. For more intense excitation pulses each medium slice instead of an isolated single-cycle wave emits some irregular and slowly decaying field oscillations, as shown in Fig.~\ref{fig8}. Besides that, the density values of the resonant atoms $N(z)$ have to be kept low enough to prevent the significant transformation of the driving pulses upon their propagation in the resonant medium, which would also cause the intense oscillating tails in the emitted field.

\section{Conclusion}

We have demonstrated that despite the common belief subcycle unipolar or quasi-unipolar pulses can be used to coherently control the induced polarization of a multi-level resonant medium. As a specific medium we considered sodium atoms (Na) and focused on 5 lowest energy levels (multiplets), namely 3S, 3P, 4S, 3D and 4P states. Under these conditions the half-cycle pulses shorter than the medium's transition periods are shown to both efficiently induce the medium polarization and to shut it down. The medium in this case emits an isolated pulse containing a few or even only one field oscillations depending on the geometry of the problem. We have also found out that for relatively weak driving pulses the response of the considered multi-level resonant medium can be well described using the two-level approximation with the ground and first excited states.

Additional degrees of freedom to tune the medium response can be provided when assuming an optically thick medium layer with a non-homogeneous spatial density distribution along the layer. As the result, both isolated half-cycle and some unusual unipolar pulses in the optical range, like rectangular- or triangular-shaped, can be produced from a medium layer. It is worth noting that the shaping of optical subcycle unipolar pulses represents a challenging issue in ultrafast optics and the paper findings can be expected to bring a considerable contribution in this area.

The obtained results make use of the control of the medium coherence (polarization) by the driving subcycle pulses, so that the coherence relaxation time $T_2$ has to largely exceed the overall duration of the excitation process. This condition can typically be readily obeyed in gaseous media, e.g. in alkali metal vapours. We thus believe that the paper findings could be experimentally validated using, for instance, plane gas jets or gas-filled hollow-core photonic-crystal fibers, in which the effectively one-dimensional propagation of ultra-short pulses can be achieved \cite{Rosanov2019}.

\section*{Acknowledgements}

Russian Science Foundation, project 21-72-10028.

% Bibliography
\bibliography{UP_library}

\end{document}